\documentclass[aps,prd,letterpaper,twocolumn,showpacs,superscriptaddress,preprintnumbers,floatfix]{revtex4-1}  
\usepackage{graphicx}  
\usepackage{dcolumn}   
\usepackage{bm}        
\usepackage{amssymb}   

\usepackage{subfigure}
\usepackage{multirow}
\usepackage{slashbox}
\usepackage{placeins}
\usepackage{morefloats}
\usepackage{epstopdf}

\usepackage{amsfonts}    
\usepackage{amsmath}
\usepackage{color}

\begin{document}

\preprint{
\vbox{
\hbox{ADP-14-3/T861}
\hbox{Edinburgh 2014/06}
\hbox{LTH 1004}
\hbox{DESY 14-016}
}}


\title{Electric form factors of the octet baryons \\from lattice QCD and chiral extrapolation}

\author{P.E.~Shanahan}\affiliation{ARC Centre of Excellence in Particle Physics at the Terascale and CSSM, School of Chemistry and Physics,
  University of Adelaide, Adelaide SA 5005, Australia}
\author{R.~Horsley}\affiliation{School of Physics and Astronomy, University of Edinburgh, Edinburgh EH9 3JZ, UK}
\author{Y.~Nakamura}\affiliation{RIKEN Advanced Institute for Computational Science, Kobe, Hyogo 650-0047, Japan}
\author{D.~Pleiter}\affiliation{JSC, Forschungzentrum J\"ulich, 52425 J\"ulich, Germany} \affiliation{Institut f\"ur Theoretische Physik, Universit\"at Regensburg, 93040 Regensburg, Germany}
\author{P.E.L.~Rakow}\affiliation{Theoretical Physics Division, Department of Mathematical Sciences, University of Liverpool, Liverpool L69 3BX, UK}
\author{G.~Schierholz}\affiliation{Deutsches Elektronen-Synchrotron DESY, 22603 Hamburg, Germany}
\author{H.~St\"uben}\affiliation{Regionales Rechenzentrum, Universit\"at Hamburg, 20146 Hamburg, Germany}
\author{A.W.~Thomas}\affiliation{ARC Centre of Excellence in Particle Physics at the Terascale and CSSM, School of Chemistry and Physics,
  University of Adelaide, Adelaide SA 5005, Australia}
\author{R.D.~Young}\affiliation{ARC Centre of Excellence in Particle Physics at the Terascale and CSSM, School of Chemistry and Physics,
  University of Adelaide, Adelaide SA 5005, Australia}
\author{J.M.~Zanotti}\affiliation{ARC Centre of Excellence in Particle Physics at the Terascale and CSSM, School of Chemistry and Physics,
  University of Adelaide, Adelaide SA 5005, Australia}

\collaboration{CSSM and QCDSF/UKQCD Collaborations}

\begin{abstract}
We apply a formalism inspired by heavy baryon chiral perturbation theory with finite-range regularization to dynamical $2+1-$flavor CSSM/QCDSF/UKQCD Collaboration lattice QCD simulation results for the electric form factors of the octet baryons. 
The electric form factor of each octet baryon is extrapolated to the physical pseudoscalar masses, after finite-volume corrections have been applied, at six fixed values of $Q^2$ in the range 0.2-1.3 GeV$^2$. The extrapolated lattice results accurately reproduce the experimental form factors of the nucleon at the physical point, indicating that omitted disconnected quark loop contributions are small relative to the uncertainties of the calculation. Furthermore, using the results of a recent lattice study of the magnetic form factors, we determine the ratio $\mu_p G^p_E/G^p_M$. This quantity decreases with $Q^2$ in a way qualitatively consistent with recent experimental results.
\end{abstract}

\pacs{12.38.Gc, 13.40.Gp, 14.20.Dh, 14.20.Jn, 12.39.Fe}
\keywords{Electric form factor, Lattice QCD, Chiral symmetry, Extrapolation}

\maketitle

\section{Introduction}
\label{sec:Introduction}

The characterization of baryon structure is a defining challenge for hadronic physics research. 
Key to such a characterization are the electromagnetic form factors which describe the spatial distribution of the charge and magnetization density in the baryons. While the nucleon form factors are precisely determined experimentally~\cite{Bernauer:2010wm,Jones:1999rz,Ron:2011rd,Zhan:2011ji,Arrington:2006zm}, those of the other octet baryons are significantly more challenging to measure and as a result are poorly known, if at all, from nature. In this light, models~\cite{Cloet:2013gva} and investigations of lattice quantum chromodynamics (QCD)~\cite{Alexandrou:2007xj,Alexandrou:2006ru,Gockeler:2003ay,Hagler:2007xi,Lin:2008uz,Liu:1994dr,Sasaki:2007gw,Hagler:2009ni,Boinepalli:2006xd,Yamazaki:2009zq,Collins:2011mk,Boinepalli:2006xd,Lin:2008mr,Wang:2008vb,MagFFs} are particularly valuable.

As the only first-principles approach which can quantitatively probe the non-perturbative domain of QCD, lattice simulations can not only provide an interpretation of experimental results for the nucleon form factors in the context of QCD but they can also give theoretical predictions of the hyperon form factors~\cite{MagFFs,Leinweber:1990dv,Boinepalli:2006xd,Lin:2008mr,Wang:2008vb}.
Furthermore, the lattice allows one to probe individual quark contributions to the form factors, giving insight into the environmental sensitivity of the distribution of quarks inside a hadron~\cite{MagFFs,Leinweber:1990dv,Boinepalli:2006xd}. 

As lattice QCD studies are limited by computation time, most simulations are performed not only at larger than physical pseudoscalar masses but often omit operator self-contractions (quark disconnected diagrams) which require noisy and expensive `all-to-all' propagators to be calculated. 
While this omission restricts the calculation of full QCD results to quantities for which disconnected contributions vanish, the comparison of experimental numbers with chirally extrapolated lattice results for other baryon observables gives insight into the significance of disconnected quark loop contributions at the physical point. This is complementary to direct lattice studies of disconnected terms~\cite{Abdel-Rehim:2013wlz,Bali:2011zzc,Alexandrou:2014eva,Babich:2010at}.

Here we analyze a subset of dynamical $2+1-$flavor CSSM/QCDSF/UKQCD Collaboration lattice simulation results for the electric form factor of the octet baryons. From these results we determine $G_E$ for all outer-ring octet baryons, at a range of discrete $Q^2$ values up to 1.3~GeV$^2$. 
We use a formalism based on connected chiral perturbation theory~\cite{Tiburzi:2009yd,Arndt:2003vd,Leinweber:2002qb,Young:2004tb} to correct for finite-volume effects and to extrapolate each baryon form factor to the physical pseudoscalar masses. The extrapolated (connected) nucleon form factors are compatible with the experimental results. This is consistent with earlier calculations of the strange form factors of the proton~\cite{Leinweber:2004tc,Leinweber:2006ug,Young:2004tb}, and with recent direct computations of disconnected contributions at larger-than-physical pion masses~\cite{Abdel-Rehim:2013wlz,Bali:2011zzc,Alexandrou:2014eva,Babich:2010at}, which suggested that disconnected effects are small.

We also supplement the lattice study of Ref.~\cite{MagFFs} by presenting new lattice simulation results for the Dirac and Pauli form factors of the outer-ring octet baryons on a larger volume and at a pion mass of 220~MeV, about 100~MeV lighter than those used in the original work. Comparison of the extrapolated (smaller volume) results with this new point provides evidence that both finite-volume effects and the chiral extrapolation are under control.

The effective field theory formalism used here is outlined in Sec.~\ref{sec:ChiPTextrap}, while the application of this formalism to the existing lattice results is presented in Sec.~\ref{sec:Fits}. Chirally extrapolated results are given in Sec.~\ref{sec:Results}, with the new lattice simulation details and a comparison of the small and large volume (light mass) results shown in Sec.~\ref{subsec:FVEffects}. Combining the determinations of the octet baryon magnetic form factors $G_M$ from Ref.~\cite{MagFFs} with this work, we present values for the ratios $\mu_B G_E^B/G_M^B$ in Sec.~\ref{sec:GEGM}.

\section{Chiral extrapolation}
\label{sec:ChiPTextrap}

To extrapolate lattice simulation results from unphysically large pseudoscalar masses to the physical point we use a formalism based on `connected chiral perturbation theory'~\cite{Tiburzi:2009yd,Arndt:2003vd,Leinweber:2002qb}, a special case of partially quenched chiral perturbation theory~\cite{Bernard:1993sv,Sharpe:2001fh,Sharpe:2000bc,Sharpe:1995qp,Chen:2001yi,Savage:2001dy,Leinweber:2002qb,Allton:2005vm}. 

Partially quenched lattice simulations traditionally employ different values for the sea and valence quark masses. As a result the distinguishing feature of the partially quenched perturbation theory formalism, developed to extrapolate such simulations, is that it allows one to treat the sea and valence quarks separately. This is precisely what is needed to extrapolate connected lattice results; the `quenching' effect is that the charges of the sea quarks are set to zero, removing the quark disconnected diagrams which are omitted from the lattice simulations. Here we use the heavy-baryon chiral perturbation theory expansion pioneered by Jenkins and Manohar~\cite{Jenkins:1990jv,Jenkins:1991ts,Jenkins:1991es,Jenkins:1991bt,Jenkins:1992pi}.

\subsection{Partially quenched chiral perturbation theory}

The nine quarks of partially quenched QCD appear in the fundamental representation of the graded symmetry group $SU(6|3)$:
\begin{equation}
\psi^T = \left( u,d,s,j,l,r,\tilde{u},\tilde{d},\tilde{s} \right).
\end{equation}
Here $(u,d,s)$ are the three usual light quarks used in hadronic interpolating fields, while $\left( \tilde{u},\tilde{d},\tilde{s}\right)$ are spin-1/2 bosonic ghost quarks. Made to be mass- and charge-degenerate with $(u,d,s)$, the ghost quarks cancel the contributions from all closed $(u,d,s)$ loops. As a result, the only disconnected loop contributions arise from the three remaining fermionic quarks $(j,l,r)$. As these quarks appear {\it only} in disconnected loops, they are interpreted as the sea-quark analogues of $(u,d,s)$.

For the extrapolation of connected lattice simulations we wish to remove disconnected quark loops entirely. This is achieved by setting the sea quark charges to zero, resulting in a quark charge matrix
\begin{equation}
\label{eq:qChMat}
Q=\textrm{diag}\left(q_u,q_d,q_s,0,0,0,q_u,q_d,q_s\right).
\end{equation}
Of course, full chiral perturbation theory is exactly reproduced by reinstating the sea quark charges through $Q \rightarrow \textrm{diag}\left(q_u,q_d,q_s,q_u,q_d,q_s,q_u,q_d,q_s\right)$~\cite{Sharpe:1995qp}.

Because the setup for this calculation parallels that of Ref.~\cite{MagFFs}, we refer to that work for further details. The next subsection presents explicit formulae for the extrapolation of $G_E$ in pseudoscalar mass.

\subsection{Electromagnetic form factors}

In the heavy-baryon formalism the electromagnetic form factors $G_E$ and $G_M$ are defined by
\begin{align}
\label{eq:jstruct}
\langle B(p') | J_\mu | B(p) \rangle = \overline{u}(p') & \left\{  \vphantom{\frac{1}{2}} \right. v_ \mu G_E(Q^2)  \\ \nonumber
& + \left. \frac{i \epsilon_{\mu \nu \alpha \beta} v^\alpha S^\beta q^\nu}{m_N} G_M(Q^2) \right\} u(p),
\end{align}
where $q=p'-p$ is the momentum transfer to the baryon $B$ and $Q^2=-q^2$. Here we focus exclusively on the electric form factor $G_E$. Expressions analogous to those in this section but for the magnetic form factor may be found in Ref.~\cite{MagFFs}.

In familiar chiral perturbation theory the leading order contribution to the electric form factor is generated by the following term in the Lagrangian:
\begin{align}
\nonumber
\mathcal{L}= -ev^\mu \left(D^\nu F_{\mu\nu}\right)  \left[ \vphantom{\left(\overline{B}\right)}\right.& b_\alpha \left( \overline{B} B Q \right) + b_\beta \left(\overline{B} Q B \right)  \\
& \left. +b_\gamma\left(\overline{B}B\right)\textrm{Str}(Q) \right].
\label{eq:elecLead}
\end{align}
For the physical quark charges the charge matrix $Q$ (Eq.~(\ref{eq:qChMat})) is such that $\textrm{Str}(Q) = 0$ and the $b_\gamma$ term does not contribute. This term is relevant only when considering individual quark contributions to the electric form factors (e.g., setting $q_u\rightarrow 1$, $q_d \rightarrow 0$, $q_s \rightarrow 0$ to obtain the $u$-quark contribution).
In line with the notation used for the magnetic form factors in Ref.~\cite{MagFFs}, we define
\begin{equation}
\label{eq:muDF}
b_\alpha = \frac{2}{3}b_D+2b_F, \hspace{3mm} b_\beta=-\frac{5}{3}b_D+b_F.
\end{equation}
Terms linear in the quark masses are generated by
\begin{align}
\label{eq:LinLag}
\nonumber
\mathcal{L}_{\textrm{lin}}=\mathcal{B}\left[ \vphantom{\left(\overline{B}^{ijk}\right)}\right.&c_1 \left(\overline{B} m_\psi B\right)\textrm{Str}(Q) + c_2\left(\overline{B}B m_\psi \right) \textrm{Str}(Q) \\ \nonumber
& + c_3 \left( \overline{B} Q B \right) \textrm{Str}(m_\psi) + c_4 \left(\overline{B} B Q \right) \textrm{Str}(m_\psi) \\ \nonumber
& + c_5 \left( \overline{B} Qm_\psi B \right)  + c_6 \left(\overline{B} B Qm_\psi \right) \\ \nonumber
& + c_7 \left(\overline{B}B \right)\textrm{Str}(Q m_\psi) + c_8 \left(\overline{B}B\right)\textrm{Str}(Q)\textrm{Str}(m_\psi) \\ \nonumber
& + c_9 (-1)^{\eta_l(\eta_j+\eta_m)}\left(\overline{B}^{kji}(m_\psi)_i^l Q_j^m B_{lmk}\right) \\ \nonumber
& + c_{10} (-1)^{\eta_j \eta_m +1} \left( \overline{B}^{kji} (m_\psi)_i^m Q_j^l B_{lmk}\right) \\ \nonumber
& + c_{11} (-1)^{\eta_l(\eta_j+\eta_m)}\left(\overline{B}^{kji}Q_i^l (m_\psi)_j^m B_{lmk}\right) \\ \nonumber
& +  \left.c_{12} (-1)^{\eta_j \eta_m +1} \left( \overline{B}^{kji}Q_i^m (m_\psi)_j^l B_{lmk}\right)\right]v^\mu \\
& \mkern-16mu \times \left(D^\nu F_{\mu\nu}\right),
\end{align}
where the shorthand for field bilinear invariants used here was originally employed by Labrenz and Sharpe in Ref.~\cite{Labrenz:1996jy}.
\begin{figure}
\begin{center}
\subfigure[]{\label{fig:mesinsoct}\includegraphics[width=0.23\textwidth]{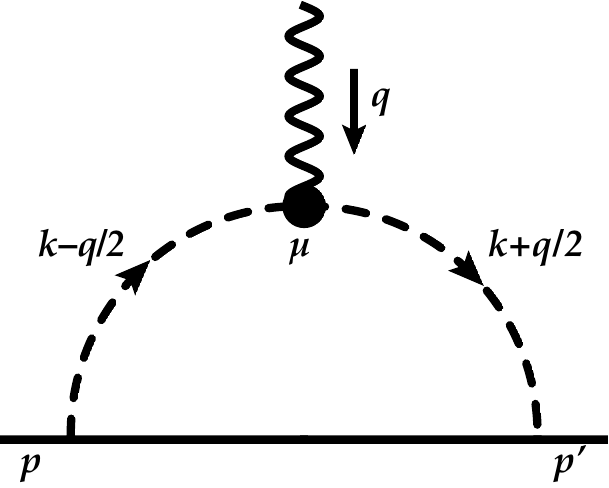}}
\subfigure[]{\label{fig:mesinsdec}\includegraphics[width=0.23\textwidth]{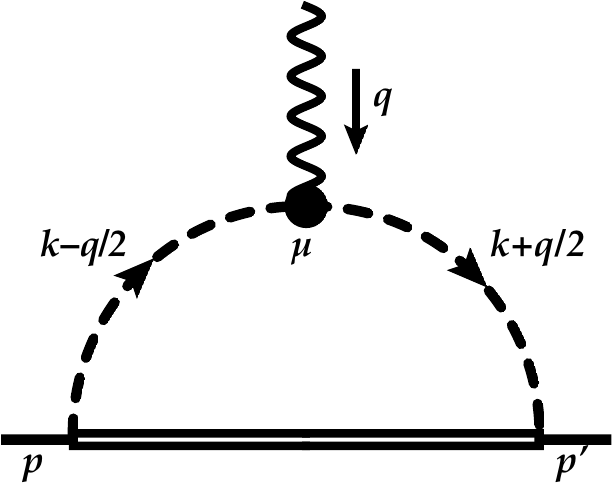}}
\subfigure[]{\label{fig:mesinstad}\includegraphics[width=0.19\textwidth]{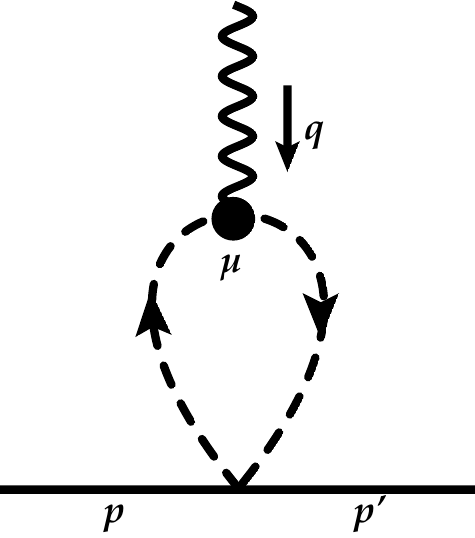}}
\caption{Loop diagrams which contribute to $G_E$ at leading order. Single, double, dashed and wavy lines represent octet baryons, decuplet baryons, mesons and photons respectively.}
\label{fig:mesinsloopst}
\end{center}
\end{figure}
The leading order loop contributions to $G_E$ are depicted in Fig.~\ref{fig:mesinsloopst}. Diagrams with both octet baryon and decuplet baryon intermediate states are included, as are tadpole loops. 

The lattice simulation results which we consider here cover values of the momentum transfer $Q^2$ up to $\approx 1.3$~GeV$^2$. This is a much larger range than can be explored with a perturbative expansion. For this reason we choose to chirally extrapolate the lattice results at fixed values of $Q^2$.
As was also done in Refs.~\cite{Wang:2008vb,Wang:2007iw,MagFFs}, we consider the coefficients in Eq.~(\ref{eq:elecLead}) to be chiral limit form factors at some \emph{fixed} $Q^2$. With a similar interpretation of the $c_i$ in Eq.~(\ref{eq:LinLag}), we can then write down chiral extrapolation formulae which have an independent set of free coefficients at each value of $Q^2$.
A particular advantage of this approach is that there is no need to impose a phenomenological constraint on the shape of the variation of the form factors with $Q^2$. Of course, a disadvantage is that we must perform independent fits to the lattice simulation results at each value of the momentum transfer.

The resulting formulae for the chiral extrapolation of the electric form factors at some \emph{fixed finite} $Q^2$ may be summarized as
\begin{align}
\nonumber
G_E^{B,q}(Q^2) =  G_E^{B,q}(Q^2=0)+Q^2\alpha^{Bq} + Q^2 \sum_{q'} \overline{\alpha}^{Bq(q')} \mathcal{B}m_{q'} \\ \nonumber
 + \frac{1}{16 \pi^3 f^2} \sum_\phi \left( \frac{1}{2} \beta^{Bq(\phi)}_{O} \mathcal{I}_{O}(m_\phi,Q^2) -\beta^{Bq(\phi)}_{D} \mathcal{I}_{D}(m_\phi,Q^2) \right. \\ \label{eq:GEExtrap}
 \left. + \beta^{Bq(\phi)}_{T} \mathcal{I}_{T}(m_\phi,Q^2) \right),
\end{align}
where $\mathcal{B}m_q$ is the mass of the quark $q$, identified with the meson masses through the appropriate Gell-Mann-Oakes-Renner relation e.g., $\mathcal{B}m_l=m_\pi^2/2$. The pion decay constant in the chiral limit is $f=0.087$~GeV~\cite{Amoros200187} (consistent with FLAG~\cite{Colangelo:2010et}) and $G_E^{B,q}(Q^2=0)$ is the total charge of the quarks of flavor $q$ in the baryon $B$. As these expressions are for quarks of unit charge, $G_E^{B,q}(Q^2=0)=2,1$ for the doubly and singly-represented quarks respectively. We point out that the parameters (e.g., $\alpha^{Bq}$) are determined independently at \emph{each} $Q^2$, so they may vary with $Q^2$.
The leading order loop contributions (Fig.~\ref{fig:mesinsloopst}) are written in terms of the integrals
\begin{align}
\label{eq:IO}
I_{O}& =  \int d\vec{k} \frac{(\vec{k}^2-\vec{q}\,^2/4)u(\vec{k}+\vec{q}/2)u(\vec{k}-\vec{q}/2)}{\omega_+\omega_-(\omega_++\omega_-)}, \\ \label{eq:ID}
I_{D}& =  \int d\vec{k} \frac{(\vec{k}^2-\vec{q}\,^2/4)u(\vec{k}+\vec{q}/2)u(\vec{k}-\vec{q}/2)}{(\omega_++\delta)(\omega_-+\delta)(\omega_++\omega_-)},\\ \label{eq:IT}
I_T & = \int d\vec{k} \frac{u(\vec{k}+\vec{q}/2)u(\vec{k}-\vec{q}/2)}{\omega_++\omega_-}
\end{align}
where $\delta$ denotes the average octet-baryon--decuplet-baryon mass splitting and
\begin{align}
\omega_+=\sqrt{(\vec{k}+\vec{q}/2)^2+m^2},\\
\omega_-=\sqrt{(\vec{k}-\vec{q}/2)^2+m^2}.
\end{align}
To prevent the charges from being renormalized by contributions from the loop integrals we make the replacement 
\begin{equation}
I(m,\vec{q}) \rightarrow \tilde{I}(m,\vec{q})=I(m,\vec{q})-I(m,0)
\label{eq:IRep}
\end{equation}
for each of the integrals above.

Within the framework of finite-range regularization, we have introduced a mass scale $\Lambda$ through a dipole regulator $u(k)=\left(\frac{\Lambda^2}{\Lambda^2+k^2}\right)^2$ inserted into the loop integrands. This shape is suggested by a comparison of the axial and induced pseudoscalar form factors of the nucleon~\cite{Guichon:1982zk}. The regulator mass is varied in the range $0.7<\Lambda<0.9$~GeV, a choice informed by a lattice analysis of nucleon magnetic moments~\cite{Hall:2012pk}. 

The finite-range regularization procedure is discussed in detail in Refs.~\cite{Leinweber:2003dg,Young:2002cj,Young:2002ib}.
Here we note that in this scheme the Goldstone boson loop contributions are suppressed at large scales by powers of $\Lambda/m_\phi$, rather than growing with powers of $m_\phi^2$. Because of this rapid suppression, all results are essentially independent of the regulator shape; replacing the dipole $u(k)$ by a monopole or Gaussian form (with appropriate ranges for $\Lambda$) yields entirely consistent results for all observables. 
We also note that higher-order terms are implicit in the structure of finite-range regularization; different regulator forms essentially correspond to different partial resummations of these terms. As a result, this scheme improves the convergence properties of the SU(3) chiral expansion and has been shown to provide a robust fit to lattice data over a large range of pion masses~\cite{Leinweber:2003dg}. 

The $\beta_T^{Bq(\phi)}$ of Eq.~(\ref{eq:GEExtrap}) are given explicitly in Appendix~\ref{app:ExtrapDetails}. The remaining coefficients, $\alpha^{Bq}$, $\overline{\alpha}^{Bq(q')}$, $\beta_O^{Bq(\phi)}$ and $\beta_T^{Bq(\phi)}$ take the same form in terms of the undetermined chiral coefficients (e.g., $c_i$) as those named identically in the case of the magnetic form factor (under the replacements $\mu_F \rightarrow b_F$ and $\mu_D \rightarrow b_D$) in Ref.~\cite{MagFFs}.
We point out that while the parameters may have the same structure for the electric and magnetic form factors, the values of the undetermined chiral coefficients are different in each case.

\section{Fits to lattice results}
\label{sec:Fits}

\begin{figure}[]
\begin{center}
\includegraphics[width=0.4\textwidth]{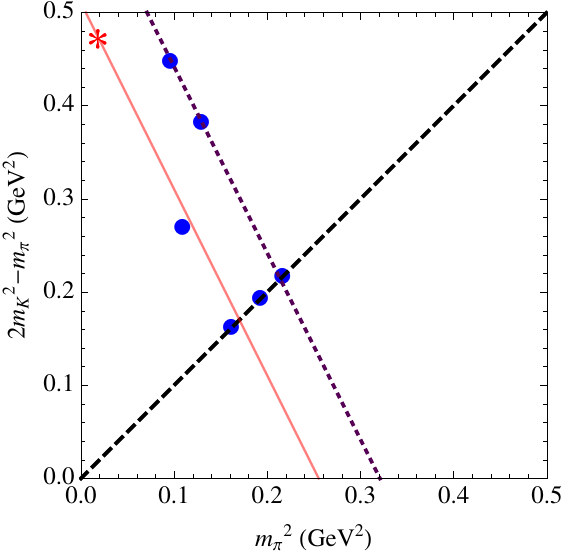}
\caption{Locations of the lattice simulation results in the $m_l-m_s$ plane. The red star denotes the physical point and the dashes indicate the flavor-symmetric line where $m_l=m_s$. Our primary simulation trajectory, illustrated by the dotted line, corresponds to the line of constant singlet quark mass $(2m_K^2+m_\pi^2)$ at $\kappa_0=0.120900$ (simulations 1--3 in Table~\ref{tab:SimDetails}). The solid line indicates the physical value of the singlet mass.}
\label{fig:mlmsPlot}
\end{center}
\end{figure}

\begin{table}[!btf]
\caption{\label{tab:SimDetails} Simulation details for the ensembles used here, with $\beta=5.50$ corresponding to $a=0.074(2)$~fm. The scale is set using various singlet quantities~\cite{Horsley:2013wqa,Bietenholz:2011qq,Bietenholz2010436}. $L^3\times T=32^3 \times 64$ for all ensembles. Raw simulation results are given in Ref.~\cite{MagFFs}.}
\begin{ruledtabular}
\begin{tabular}{ccccccc}
 & $\kappa_0$ & $\kappa_l$ & $\kappa_s$  & $m_\pi$~(MeV) & $m_K$~(MeV) & $m_\pi L$ \\
\hline
1 & 0.120900 & 0.120900 & 0.120900 &   465 & 465 & 5.6 \\
2 &  & 0.121040 & 0.120620 & 360 & 505 & 4.3 \\
3 &  & 0.121095 & 0.120512 & 310 & 520 & 3.7 \\ \hline
4 & 0.120920 & 0.120920 & 0.120920 & 440 & 440 & 5.3 \\ \hline
5 & 0.120950 & 0.120950 & 0.120950 & 400 & 400 & 4.8 \\
6 &  & 0.121040 & 0.120770 & 330 & 435 & 4.0  \\
\end{tabular}
\end{ruledtabular}
\end{table}
\vspace{-10pt}

The CSSM/QCDSF/UKQCD lattice simulation results which we use for this study, summarized in Fig.~\ref{fig:mlmsPlot} and Table~\ref{tab:SimDetails}, were presented in terms of the Dirac and Pauli form factors $F_1$ and $F_2$ in Ref.~\cite{MagFFs}. Here we consider the electric Sachs form factor $G_E$ which may be obtained as the linear combination:
\begin{align}
G_E(Q^2)&=F_1(Q^2)-\frac{Q^2}{4M_N^2}F_2(Q^2).
\end{align}

Before fitting the chiral perturbation theory expressions of Sec.~\ref{sec:ChiPTextrap} to the lattice simulation results, we correct the raw lattice data for small finite volume effects. This procedure is explained in Refs.~\cite{Hall:2012yx,Hall:2013oga,MagFFs}, and involves shifting the lattice points by the difference found by replacing the infinite-volume integrals of the leading-order chiral loop integral expressions with finite-volume sums. 
As momentum is quantized on the lattice, the finite-volume sums must be calculated with the integrands in Eqs.~(\ref{eq:IO}), (\ref{eq:ID}) and (\ref{eq:IT}) shifted from being symmetric (meson lines with momenta $k-q/2$ and $k+q/2$) to what is more natural for the lattice, namely meson lines with momenta $k$ and $k+q$.

One possible artifact in this estimate of the finite-volume corrections is that the naive enforcement of charge-nonrenormalization by Eq.~(\ref{eq:IRep}) may lead to an overestimate of the corrections at large values of the momentum transfer $Q^2$.
While the higher-order diagrams (not included here) which would naturally prevent the renormalization of charge would contribute less at large $Q^2$, the constant subtraction used here does not have that feature. As the finite-volume corrections are nevertheless small -- neglecting them yields results for all relevant observables which are consistent within uncertainties with those presented here -- this is not a significant effect.

\begin{figure}
\begin{center}
\includegraphics[width=0.48\textwidth]{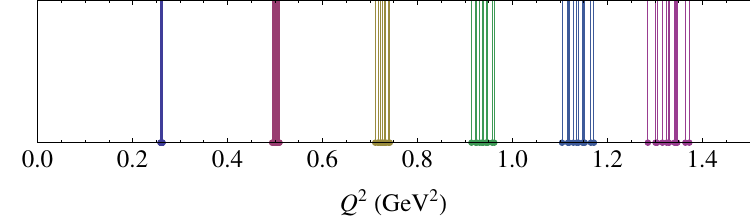}
\caption{$Q^2$ distribution for the lattice simulation results. Colors indicate the $Q^2$ bin groupings; each bin corresponds to a single value of the three-momentum transfer in lattice units.}
\label{fig:Q2BinPlots}
\end{center}
\end{figure}

\begin{figure}
\begin{center}
\includegraphics[width=0.48\textwidth]{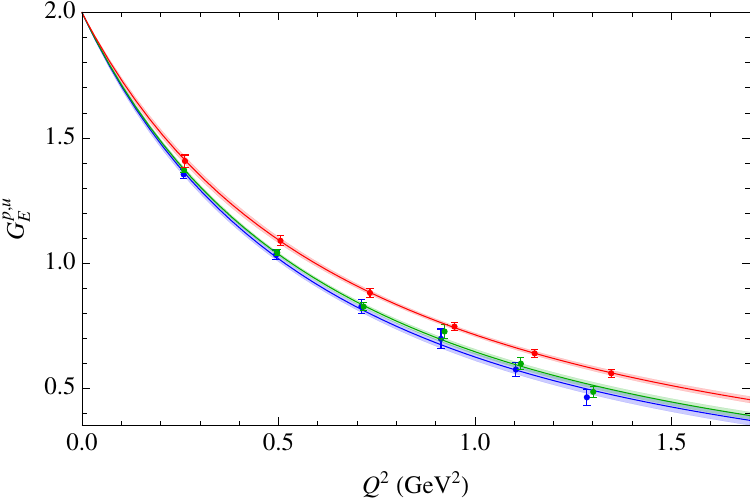}
\caption{Generalized dipole fits (Eq.~\ref{eq:gendipfit}) upon which the binning corrections are based. The three fits shown correspond to the three different pseudoscalar mass points along the primary simulation trajectory (simulations 1-3 in Table~\ref{tab:SimDetails}). Quarks have unit charge.}
\label{fig:DipoleFits}
\end{center}
\end{figure}

The chiral extrapolation expressions of Sec.~\ref{sec:ChiPTextrap} are derived for fixed values of the momentum transfer $Q^2$. For this reason, we perform six independent fits to the lattice simulation results; one fit to each bin of data corresponding to a single value of $Q^2$ in lattice units. As the physical values of $Q^2$ in each bin vary slightly because of the range of pseudoscalar and baryon masses considered, illustrated in Fig.~\ref{fig:Q2BinPlots} (the largest variation is in the range $1.29-1.37$~GeV$^2$ for the highest Q$^2$ bin), all simulation results are shifted to the average $Q^2$ value of their respective bin. This shift is performed using a dipole-like fit to the (finite-volume--corrected) simulation results.
The functional form used is
\begin{equation}
\label{eq:gendipfit}
G^\textrm{fit}_E(Q^2)=\frac{G_E(Q^2=0)}{1+d_1Q^2+d_2Q^4},
\end{equation}
where $d_1$ and $d_2$ are free parameters and $G_E(Q^2=0)=1,2$ for the singly and doubly represented quarks (of unit charge) respectively. This particular functional form is chosen as it provides a good fit to the lattice simulation results; as illustrated later, a standard dipole form performs poorly. Several examples of the fits are shown in Fig.~\ref{fig:DipoleFits}.

\begin{figure}
\begin{center}
\subfigure[Lowest $Q^2$ bin: $Q^2 \approx 0.26$~GeV$^2$.]{
\includegraphics[width=0.48\textwidth]{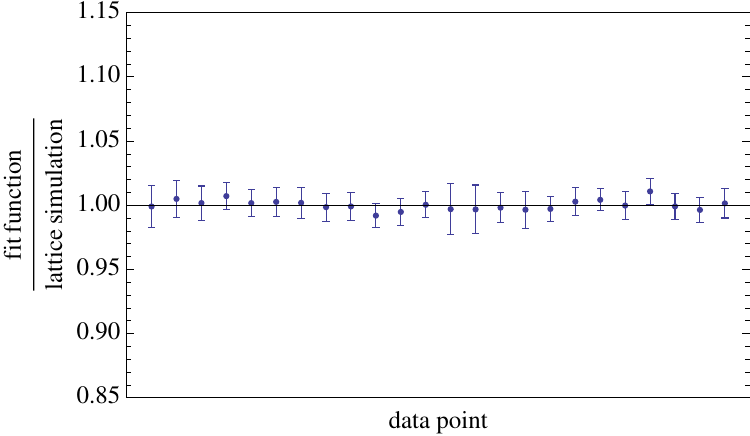} \label{subfig:1}
}
\subfigure[Highest $Q^2$ bin: $Q^2 \approx 1.35$~GeV$^2$.]{
\includegraphics[width=0.48\textwidth]{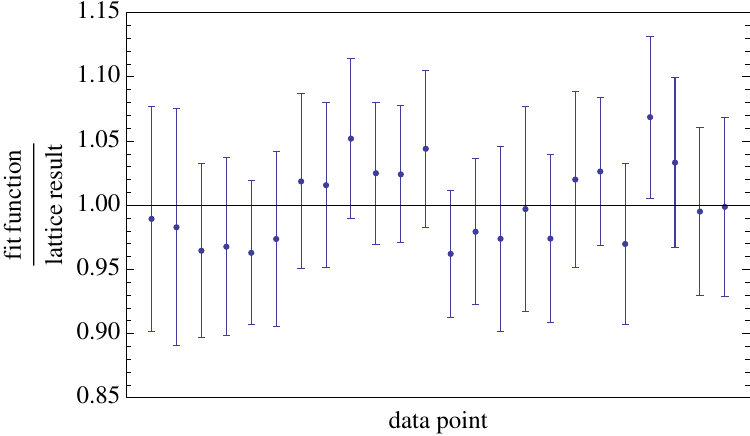} \label{subfig:2}
}
\caption{Illustration of the quality of fit for the lowest and highest $Q^2$ bins. Each point denotes one of the lattice simulation results e.g., $G_E^{p,u}$, $G_E^{p,d}$ \ldots, at one of the sets of pseudoscalar masses of Table~\ref{tab:SimDetails}. The comparison of Figs.~\ref{subfig:1} and \ref{subfig:2} shows the expected increase in uncertainty as $Q^2$ increases (i.e., as one moves further from $Q^2=0$ where the value of $G_E$ is fixed).}
\label{fig:FitQual}
\end{center}
\end{figure}

After the lattice simulation results have been finite-volume corrected and binned in $Q^2$, we perform an independent fit, using Eq.~(\ref{eq:GEExtrap}), to the variation with $m_\pi$ and $m_K$ of the results in each bin. This involves a simultaneous fit, at the bootstrap level, to all octet baryon form factors ($G_E^{p,u}$, $G_E^{p,d}$, $G_E^{\Sigma,u}$, $G_E^{\Sigma,s}$, $G_E^{\Xi,s}$ and $G_E^{\Xi,u}$) at each of the six sets of pseudoscalar masses of Table~\ref{tab:SimDetails}. There are 24 data points (6 at each of the points for which $m_\pi\ne m_K$ and 2 at each SU(3)-symmetric point), and 8 fit parameters, at each $Q^2$. 
Figure~\ref{fig:FitQual} illustrates the fit quality for the highest and lowest $Q^2$ bins, which are representative of all six fits.
Values of the fit parameters, which are the undetermined chiral coefficients $b_{D/F}$ and relevant linear combinations of the $c_{i}$, $c_{ij}$, are shown in Appendix~\ref{app:fitParams}.

As the fits are performed using an adaptation of connected chiral perturbation theory applied to connected lattice simulation results, they yield closed-form functions for the connected contribution to the octet baryon electric form factors as a function of pion and kaon mass, at each simulation $Q^2$.

\begin{figure}
\begin{center}
\includegraphics[width=0.48\textwidth]{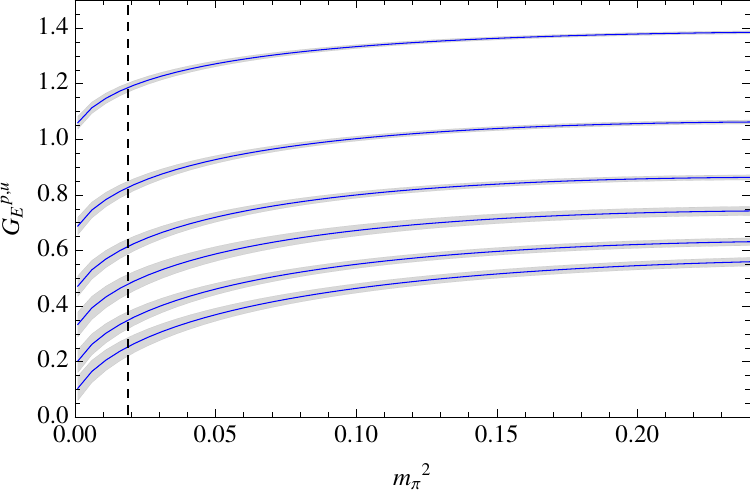}
\caption{Up quark (connected) contribution to the proton electric form factor for quarks with unit charge. Each set of results (top to bottom) represents an independent fit at a different (increasing) value of $Q^2$. The lines show these fits evaluated along the trajectory which holds the singlet pseudoscalar mass $(m_K^2+m_\pi^2/2)$ fixed to its physical value.}
\label{fig:TotalPuPlot}
\end{center}
\end{figure}

\section{Analysis of results}
\label{sec:Results}

In this section we present and discuss the results of the chiral extrapolations. Figure~\ref{fig:TotalPuPlot} shows the extrapolated up quark (connected) contribution to the proton electric form factor for all 6 values of the momentum transfer $Q^2$. The trajectory chosen illustrates the variation of the form factor with $m_\pi^2$ at fixed (physical) singlet mass $(m_K^2+m_\pi^2/2)$.

The following subsections present chirally extrapolated results at the physical pseudoscalar masses for some observables of interest: isovector quantities (Sec.~\ref{sec:isovector}), connected form factors  which give insight into the magnitude of disconnected terms (Sec.~\ref{sec:connected}), electric radii (Sec.~\ref{sec:radii}) and quark form factors which allow one to investigate the environmental sensitivity of the distribution of quarks inside a baryon (Sec.~\ref{subsec:quarkFFs}). Finally, the results of a new lattice simulation, at a lighter pion mass and larger volume than the primary set of results considered here, are presented in Sec.~\ref{subsec:FVEffects}. A comparison of the extrapolated smaller volume results with these new numbers allows one to gauge the extent to which finite volume and pion mass effects are under control in this study.   

\subsection{Isovector quantities}
\label{sec:isovector}

Isovector combinations of observables are of particular interest to this study as they can be determined from connected lattice results with the smallest systematic uncertainty. As disconnected quark loops, which are omitted from the lattice simulations and extrapolations, cancel for these combinations, the extrapolated results may be directly compared with experimental numbers.

\begin{figure}[]
\begin{center}
\includegraphics[width=0.48\textwidth]{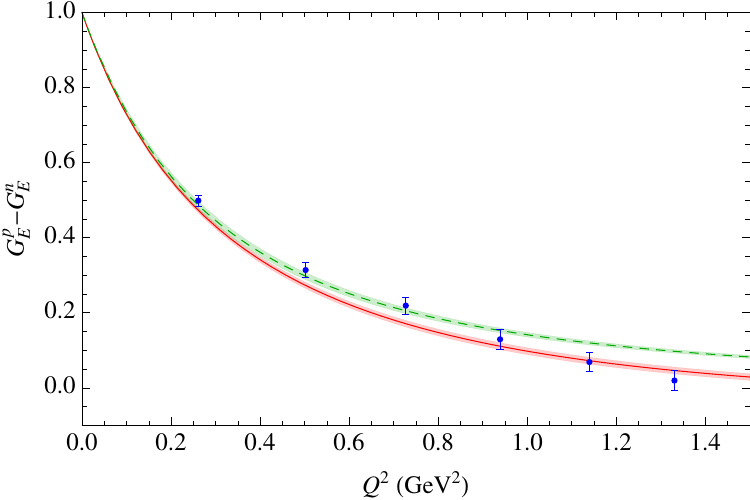}
\caption{Isovector nucleon electric form factor compared to the Kelly parameterization of experimental results~\cite{Kelly:2004hm} (red solid line). The failure of a simple dipole fit to the simulation results to provide a satisfactory description is illustrated by the green (dashed) line.}
\label{fig:KellyIsovec}
\end{center}
\end{figure}

\begin{figure}[]
\begin{center}
\subfigure[Isovector sigma baryon electric form factor.]{
\includegraphics[width=0.48\textwidth]{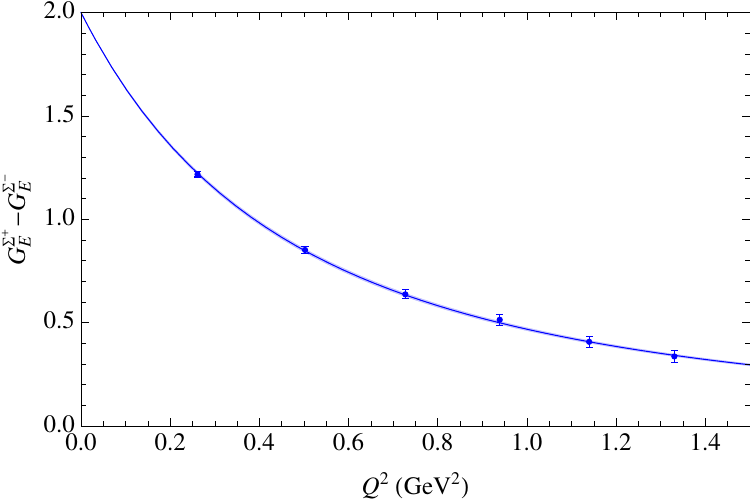}
\label{fig:IsoSig}
}
\subfigure[Isovector cascade baryon electric form factor.]{
\includegraphics[width=0.48\textwidth]{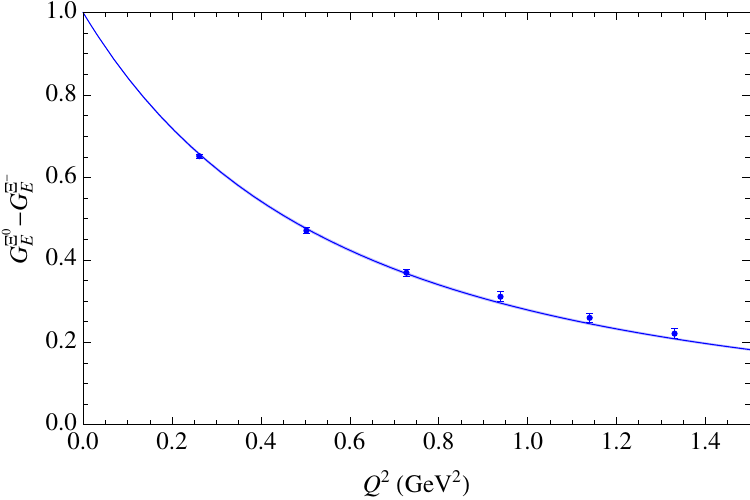}
\label{fig:IsoXi}
}
\caption{Isovector sigma and cascade baryon electric form factors with dipole-like fits in $Q^2$ (Eq.(\ref{eq:diplike})).}
\label{fig:IsoSigXi}
\end{center}
\end{figure}

Figure~\ref{fig:KellyIsovec} shows the impressive comparison of the extrapolated isovector nucleon form factor with the Kelly parameterization of experimental results~\cite{Kelly:2004hm}. The agreement is remarkable across the entire range of $Q^2$ values considered. We note, however, that a dipole form, also illustrated in Fig.~\ref{fig:KellyIsovec}, does not provide a good fit to the extrapolated results, with the $\chi^2/\textrm{d.o.f}\approx 3.2$.  A more general dipole-like fit function:
\begin{equation}
G^\textrm{fit}_E(Q^2)=\frac{G_E(Q^2=0)}{1+d_1Q^2+d_2Q^4+d_3Q^6},
\label{eq:diplike}
\end{equation}
performs significantly better, with the $\chi^2/\textrm{d.o.f}\approx 1$.
As our previous study~\cite{MagFFs} indicates that $G_M$ is described acceptably by a dipole form in $Q^2$, this suggests that $G_E/G_M \neq$ constant. This is discussed further in Sec.~\ref{sec:GEGM}.

The isovector combinations of sigma and cascade baryon electric form factors are shown in Fig.~\ref{fig:IsoSigXi}. As no experimental results are available for these form factors, dipole-like fits (Eq.~(\ref{eq:diplike})) to the extrapolated simulation results have been included to guide the eye.

\begin{figure}[]
\begin{center}
\subfigure[Proton electric form factor.]{
\includegraphics[width=0.48\textwidth]{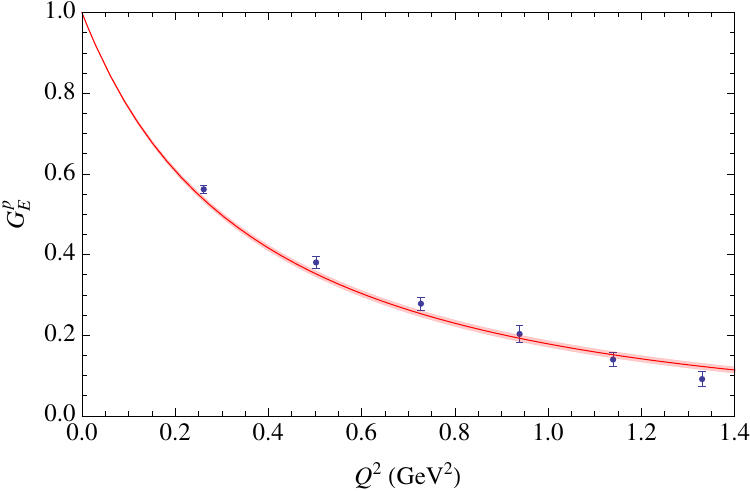}
\label{fig:pKelly}
}
\subfigure[Neutron electric form factor.]{
\includegraphics[width=0.48\textwidth]{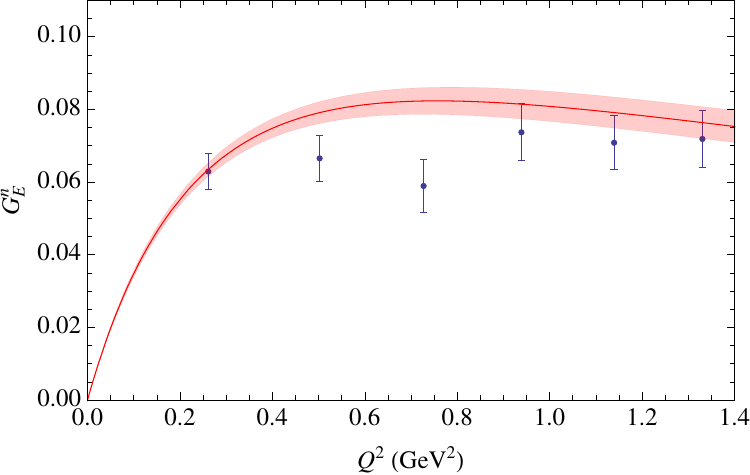}
\label{fig:nKelly}
}
\caption{Extrapolated (connected part of the) proton and neutron electric form factors, compared with Kelly parameterization~\cite{Kelly:2004hm} of experimental measurements.}
\label{fig:KellyComp}
\end{center}
\end{figure}

\subsection{Connected quantities}
\label{sec:connected}

The excellent agreement of the extrapolated isovector nucleon form factor with the experimental result suggests that this lattice study may provide a good indication of the significance of the omitted disconnected quark loop contributions to the form factors.
In particular, we compare the `connected part' of the proton and neutron electric form factors with the experimental values. Any deviation from experiment more significant than that of the isovector results could indicate, among other systematic effects, an important contribution from disconnected loops.
Figure~\ref{fig:KellyComp} shows extrapolated results for the connected parts of the proton and neutron electric form factors, compared with the Kelly parameterization of experimental results~\cite{Kelly:2004hm}. The outstanding agreement between the lattice and experimental results, at all values of $Q^2$, indicates that the omitted disconnected contributions are small compared with the uncertainties of this calculation, provided that other systematic effects, such as excited-state contamination, are negligible. This is consistent with the results of recent direct lattice studies of disconnected quantities at larger values of the pion mass~\cite{Abdel-Rehim:2013wlz,Bali:2011zzc}. 

We note that as only one value of the source-sink separation is used here~\cite{MagFFs}, it is difficult to estimate the size of excited-state contamination effects quantitatively away from the $Q^2=0$ limit where they must vanish. A detailed study similar to that of Ref.~\cite{Green:2014ez} 
would be a valuable extension of this analysis.

Figures displaying results for each of the remaining outer-ring octet baryons, including dipole-like (Eq.~(\ref{eq:diplike})) fits in $Q^2$ for the charged baryons, are given in Appendix~\ref{app:FFPics}.

\subsection{Electric radii}
\label{sec:radii}

The electric radii of the charged octet baryons are defined by
\begin{equation}
\langle r^2_E \rangle^B = -\frac{6}{G_E(Q^2=0)}\frac{d}{dQ^2}G_E^B(Q^2)\bigg|_{Q^2=0}. 
\end{equation}
To evaluate this expression from the lattice simulation results we first extrapolate the electric form factors to the physical pseudoscalar masses, at each simulation $Q^2$ value, as described in the previous section. The extraction of the electric radii is then performed by fitting some form to the variation in $Q^2$ of those extrapolated results. We consider here both a traditional dipole and a more general dipole-like (Eq.~(\ref{eq:diplike})) ansatz. As was noted previously for the isovector nucleon form factor, the dipole form does not provide a good fit to the extrapolated lattice results; the $\chi^2/\textrm{dof}$ is as large as 4.0 for the $\Xi^-$ and 1.7 for the proton. In contrast, the more general form of Eq.~(\ref{eq:diplike}) yields fits with a $\chi^2/\textrm{dof} \lesssim 1$ for each of the charged baryons.  
Fits using this ansatz are shown in Appendix~\ref{app:FFPics}. 
Results for the radii of the charged baryons, compared with the available experimental numbers, are given in Table~\ref{tab:ElecRadii}.

\begin{table}[]
\caption{\label{tab:ElecRadii} Octet baryon electric radii based on a dipole or dipole-like (Eq.~(\ref{eq:diplike})) fit to the extrapolated lattice simulation results, compared with the experimental values~\cite{PDG}. }
\begin{ruledtabular}
\begin{tabular}{lD{.}{.}{5}D{.}{.}{4}D{.}{.}{4}D{.}{.}{4}}
 & \multicolumn{4}{c}{$\langle r^2_E \rangle^B$~(fm$^2$)}\\
 & \multicolumn{1}{c}{$p$} &   \multicolumn{1}{c}{$\Sigma^+$} &  \multicolumn{1}{c}{$\Sigma^-$} &   \multicolumn{1}{c}{$\Xi^-$} \\
\hline
Dipole ansatz & 0.601(14) & 0.598(12) & 0.414(5) & 0.352(3) \\
Eq.~\ref{eq:diplike} ansatz & 0.76(10) & 0.61(8) & 0.45(3) & 0.37(2)  \\
Experimental & 0.878(5) &  & 0.780(10) &   \\
\end{tabular}
\end{ruledtabular}
\end{table}

The electric radii determined by this method are consistently smaller than the corresponding experimental numbers for the proton and $\Sigma^-$. We point out that while this calculation omits any disconnected contributions to the form factors and therefore to the radii, the very close agreement of the extracted proton electric form factor with the experimental determination suggests that the effect of this omission is small, barring lattice artefacts as discussed in the previous section.
It is clear that the simple dipole-like parameterization used for the $Q^2$-dependence is not sufficient to extract accurate values of the electric radii from these simulations. Robust predictions of the electric radii from lattice QCD will require simulations with a similar level of precision to the results of this work, but at much lower $Q^2$ values.

\begin{figure}
\includegraphics[width=0.48\textwidth]{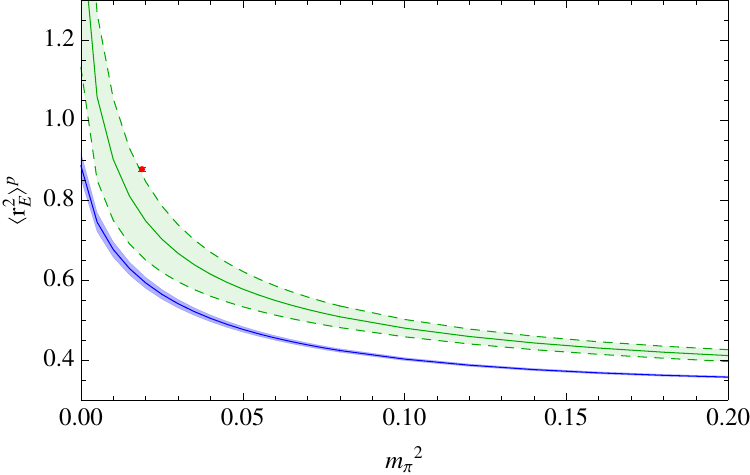}
\caption{Electric radius of the proton from the chiral extrapolation, with a dipole (blue band) or dipole-like (green dashed band) ansatz (Eq.~(\ref{eq:diplike})) parameterizing the $Q^2$-dependence. The singlet pseudoscalar mass $(m_K^2+m_\pi^2/2)$ is held fixed to its physical value. The red point indicates the experimental value.}
\label{fig:ProtonRad}
\end{figure}

We note that the electric radius of the proton extracted as described above does display the expected behaviour with pion mass, increasing quite rapidly as one approaches the physical pseudoscalar masses from above. This is illustrated in Fig.~\ref{fig:ProtonRad}. 

\subsection{Quark form factors}
\label{subsec:quarkFFs}

\begin{figure}[]
\begin{center}
\subfigure[Doubly-represented quark contributions.]{
\includegraphics[width=0.48\textwidth]{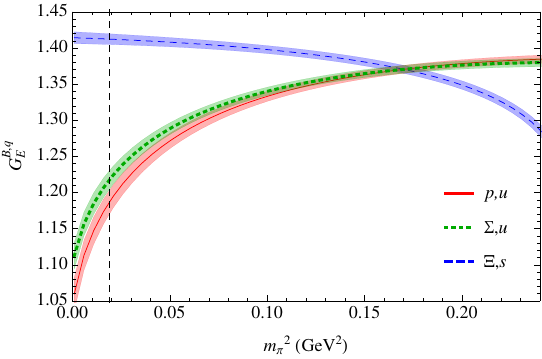}
\label{fig:DoublyRep}
}
\subfigure[Singly-represented quark contributions.]{
\includegraphics[width=0.48\textwidth]{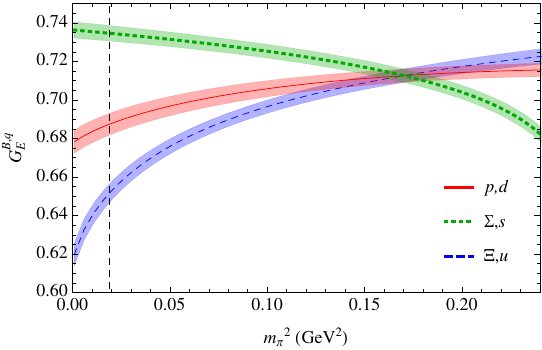}
\label{fig:SinglyRep}
}
\caption{Connected part of the doubly and singly-represented quark contributions to the baryon electric form factors for $Q^2\sim 0.26$~GeV$^2$. The charges of the relevant quarks have been set to one and the singlet mass $(m_K^2+m_\pi^2/2)$ is held fixed to its physical value.}
\label{fig:quarkreps}
\end{center}
\end{figure}

We investigate the environmental sensitivity of the distribution of quarks inside a hadron by inspecting the individual (connected) quark contributions to the electric form factors of the octet baryons. These contributions, evaluated using the chiral extrapolation described in previous sections, are illustrated in Fig.~\ref{fig:quarkreps}. The figures show the lowest $Q^2$ result, at approximately 0.26~GeV$^2$. We recall that the lines shown on each plot are not independent as the chiral extrapolation expressions are simultaneously fit to all of the octet baryon form factors.

The doubly-represented quark contributions to the form factors are illustrated in Fig.~\ref{fig:DoublyRep}. While the $u$ contribution to the proton and the $u$ contribution to the sigma baryon are very similar -- the only difference is the mass of the single spectator ($d$ or $s$) quark -- the $s$ contribution to the cascade baryon has a different shape. That form factor has significantly less curvature with $m_\pi^2$ below the SU(3)-symmetric point as a result of the heavier mass of the probed $s$ quark.  

The singly-represented quark contributions are shown in Fig.~\ref{fig:SinglyRep}. Here the difference between the $d$ quark contribution to the proton and the $s$ quark contribution to the sigma baryon illustrates the effect of changing the mass of the single probed quark. 
While the effect of changing the mass of the spectator quark is small for the doubly-represented form factors, it is far more significant here as there are now two spectator quarks. This may be seen by comparing the $d$ quark contribution to the proton with the $u$ in the cascade baryon.

We notice that the $u$ quark contribution to the cascade baryon is considerably more suppressed in the light quark-mass region than the corresponding $d$ quark contribution to the proton. That is, the magnitude of $\langle r^2\rangle_u^\Xi$ is enhanced relative to $\langle r^2\rangle_d^p$. This can be explained by the meson-dressing effects; the connected $d$ in the proton prefers to form a $\pi^+$ with one of the valence $u$ quarks in the proton, giving rise to a substantial negative contribution to $\langle r^2\rangle_d^p$ in the light quark-mass region. In contrast, the connected $u$ in the cascade baryon can only form a pion state by coupling to a sea quark, from which the resulting enhancement is always positive.

\subsection{Finite-volume effects}
\label{subsec:FVEffects}

\begin{figure*}[!h]
\begin{center}
\subfigure[]{
\includegraphics[width=0.48\textwidth]{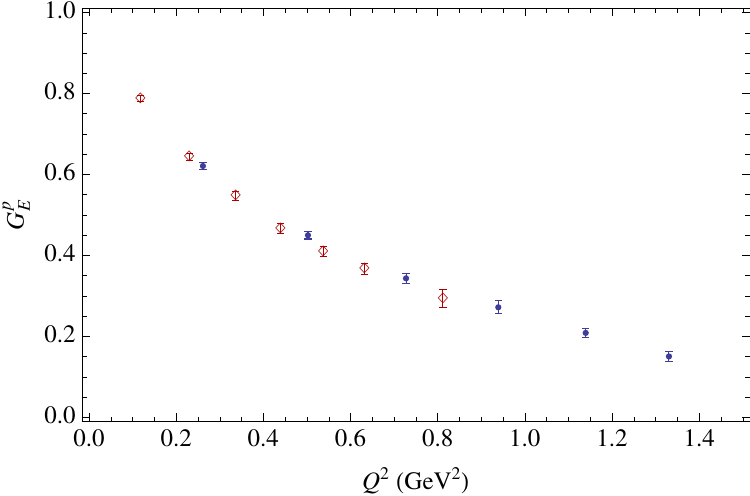}
}
\subfigure[]{
\includegraphics[width=0.48\textwidth]{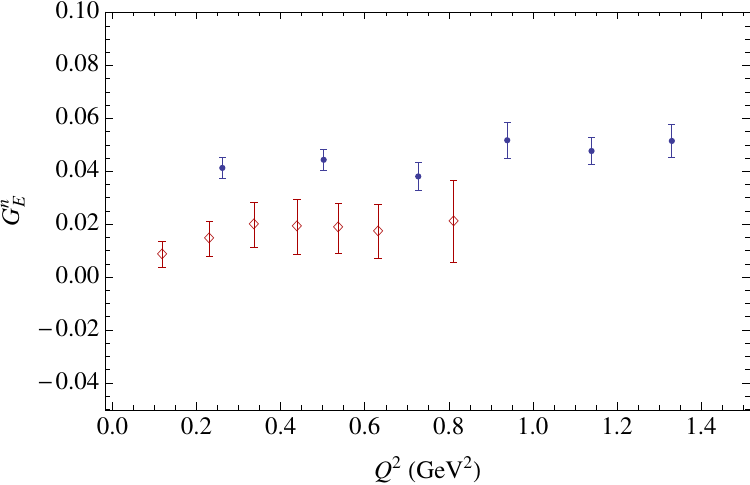}
}
\subfigure[]{
\includegraphics[width=0.48\textwidth]{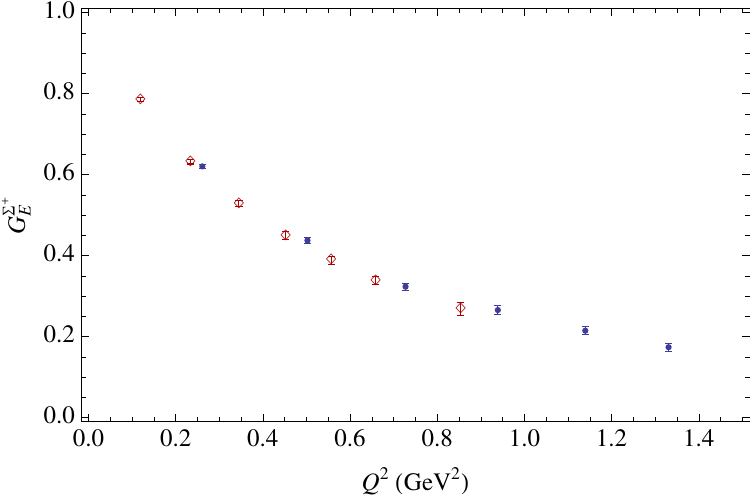}
}
\subfigure[]{
\includegraphics[width=0.48\textwidth]{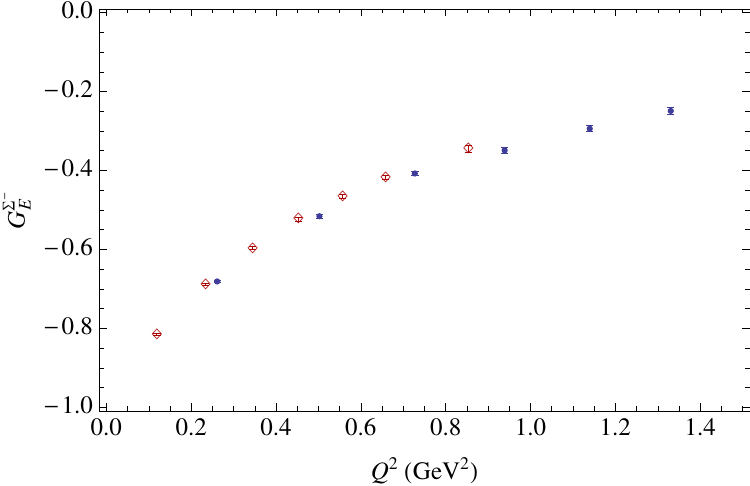}
}
\subfigure[]{
\includegraphics[width=0.48\textwidth]{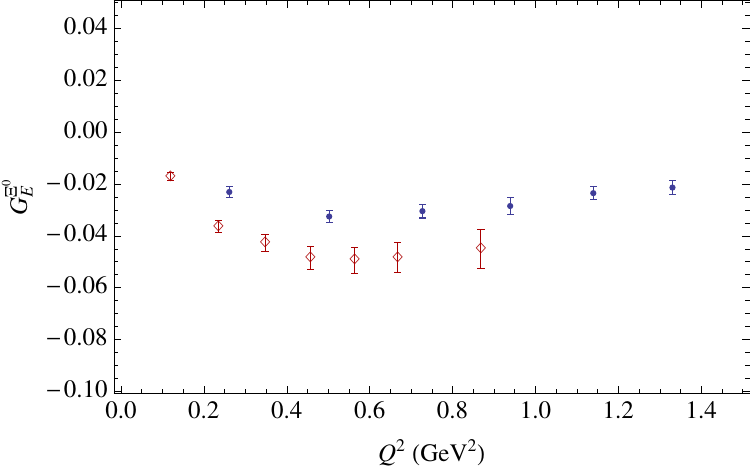}
}
\subfigure[]{
\includegraphics[width=0.48\textwidth]{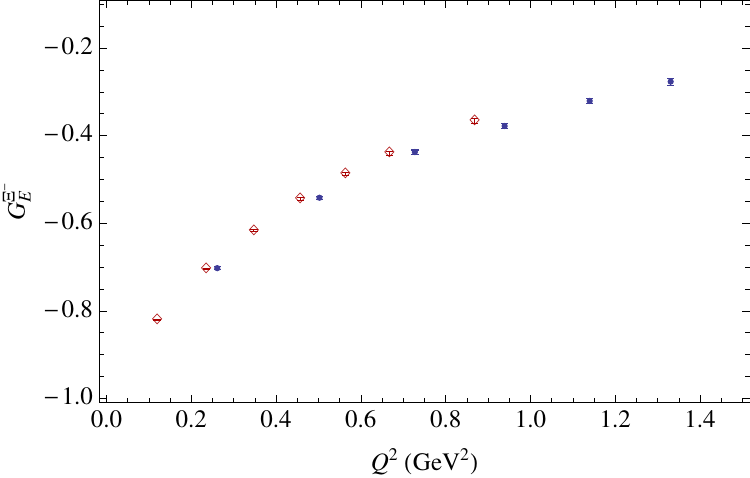}
}
\caption{Connected part of the octet baryon electric form factors at the pseudoscalar masses of simulation 7 in Table~\ref{tab:SimDetailsNew}, $(m_\pi,m_K)=(220,540)$~MeV. Solid blue circles indicate the results of the chiral extrapolation of the $32^3\times 64$ volume lattice simulation results to these masses, while the empty red diamonds indicate the $48^3\times 96$ volume results without any extrapolation. Finite volume corrections, based on leading order perturbation theory, have been applied to all results. }
\label{fig:FVComp}
\end{center}
\end{figure*}

One limitation of the analysis presented in the previous sections is that all of the lattice simulations were performed on a single $32^3\times 64$ volume. Although finite-volume corrections, based on leading order chiral perturbation theory, have been performed, it is instructive to check that finite volume effects have indeed been properly accounted for by comparing to lattice simulation results on larger volumes.
To facilitate this an additional simulation has been performed on a larger $48^3\times 96$ volume at a lighter pion mass $m_\pi=220$~MeV. Simulation details for this new ensemble are given in Table~\ref{tab:SimDetailsNew}. The lattice set-up is entirely analogous to that of the other simulations considered in this study; the gauge field configurations have been generated with $N_f=2+1$ flavors of dynamical fermions using the tree-level Symanzik improved gluon action and nonperturbatively $\mathcal{O}(a)$ improved Wilson fermions. We refer to Refs.~\cite{Bietenholz:2011qq,Bietenholz2010436} for further details. Raw lattice results for $F_1$ and $F_2$ are given in Appendix~\ref{app:RawLatt}. 

As there is only one new simulation on the larger volume, and the discrete $Q^2$ values in physical units differ substantially between volumes, we do not include this new simulation into the chiral perturbation theory fits. 
Instead we compare the results of the fits, extrapolated \emph{to the pseudoscalar masses of the new point} (with a pion mass about 100~MeV lighter than the lightest pion mass of ensembles 1-6 in Table~\ref{tab:SimDetails}), with the larger-volume results. We note that finite-volume corrections, as outlined in Sec.~\ref{sec:Fits}, have been applied to the new results.

Figure~\ref{fig:FVComp} shows the excellent agreement between the chirally extrapolated small-volume results and the larger-volume results for the charged baryons. 
For the neutral form factors in particular there is a systematic shift between the results on the two volumes, although we point out that the absolute magnitude of this shift is small -- of the order of $5\%$ of the proton form factor. This is comparable to the size of the discrepancies between the charged baryon form factors on the two volumes.
The shift may be evidence of excited state contamination in either set of results -- which can not be estimated quantitatively as only one value of the source-sink separation is used here -- or the effect of some other yet-to-be-understood systematic.
Nevertheless, the comparison is extremely encouraging and suggests that both the systematic finite-volume effect and the extrapolation in pion mass are well under control for the charged baryon form factors.

\begin{table}[!btf]
\caption{\label{tab:SimDetailsNew} Simulation details for the new ensemble, with $L^3\times T=48^3 \times 96$ and $\beta=5.50$ corresponding to $a=0.074(2)$~fm. The scale is set using various singlet quantities~\cite{Horsley:2013wqa,Bietenholz:2011qq,Bietenholz2010436}. Raw simulation results for $F_1$ and $F_2$ are given in Appendix~\ref{app:RawLatt}.}
\begin{ruledtabular}
\begin{tabular}{cccccc}
 & $\kappa_l$ & $\kappa_s$  & $m_\pi$~(MeV) & $m_K$~(MeV) & $m_\pi L$ \\
\hline
7 & 0.121166  & 0.120371 &   220 & 540 & 4.0 \\
\end{tabular}
\end{ruledtabular}
\end{table}

\section{Ratio of electric and magnetic form factors}
\label{sec:GEGM}

\begin{figure}[]
\begin{center}
\includegraphics[width=0.48\textwidth]{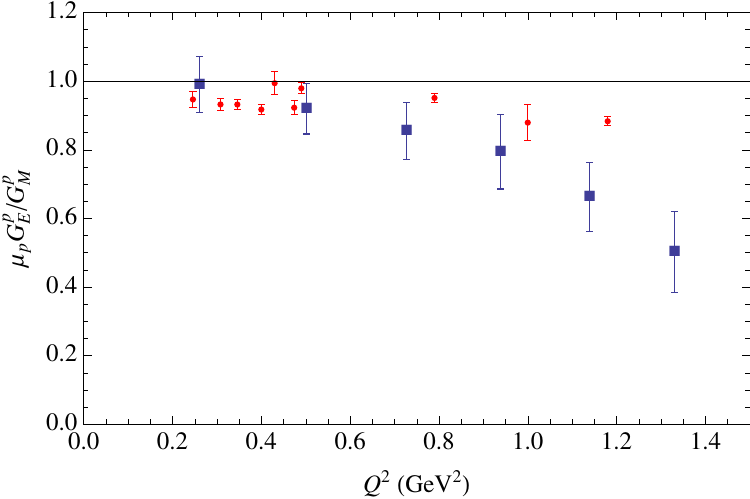}
\caption{The blue squares show the ratios of the electric (from this work) and magnetic (from Ref.~\cite{MagFFs}) form factors for the proton. The red circles denote the experimental results of Refs.~\cite{PhysRevC.84.055204,PhysRevC.73.064004,PhysRevC.71.055202}. }
\label{fig:RatiosP}
\end{center}
\end{figure}

\begin{figure}[]
\begin{center}
\subfigure[Charged baryons.]{
\includegraphics[width=0.48\textwidth]{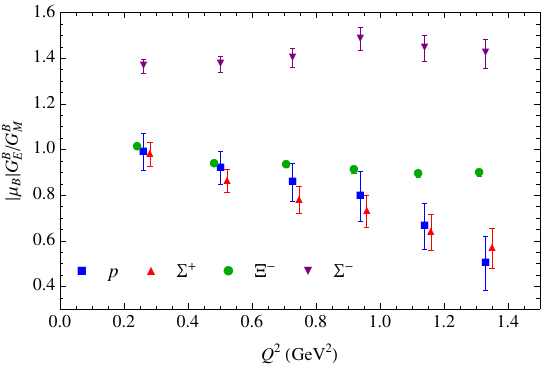}
\label{fig:chargedratios}
}
\subfigure[Neutral baryons.]{
\includegraphics[width=0.48\textwidth]{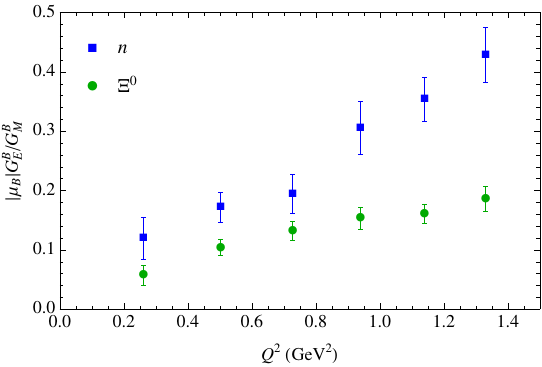}
\label{fig:neutralratios}
}
\caption{Ratios of the electric (from this work) and magnetic (from Ref.~\cite{MagFFs}) form factors of the octet baryons. The points denoting the $\Sigma^+$ and $\Xi^-$ baryons in Fig.~\ref{fig:chargedratios} have been slightly offset on the $Q^2$ axis for clarity. }
\label{fig:Ratios}
\end{center}
\end{figure}

By combining the chirally extrapolated values of $G_E$ from this work with the magnetic form factors $G_M$ determined in Ref.~\cite{MagFFs}, we are able to deduce the ratio $\mu G_E/G_M$ at each of the six discrete values of $Q^2$ for which we have results. The largest $Q^2$ value is $\approx 1.3$~GeV$^2$.
As both this work and Ref.~\cite{MagFFs} use the same extrapolation techniques and are based on the same set of CSSM/UKQCD/QCDSF lattice simulation results, the analysis can be done at the bootstrap level.

Figure~\ref{fig:RatiosP} shows the proton form factor ratio $\mu_p G^p_E/G^p_M$, where the experimental value is used for the magnetic moment $\mu_p$~\cite{PDG}. While the results are qualitatively consistent with a linear decrease of the ratio with $Q^2$ as concluded from polarization transfer experiments (e.g., from Refs.~\cite{PhysRevC.84.055204,PhysRevC.73.064004,PhysRevC.71.055202}, illustrated on the figure), this decrease is more pronounced in our results than in the experimental data with the exception of the results of Ref.~\cite{PhysRevLett.105.242001} which display a similarly steep trend. In our work this trend is explained by the observation~\cite{MagFFs} that the lattice simulation results for $G_M$ fall off less rapidly in $Q^2$ than the experimental results, while the lattice results for $G_E$ are consistent with experiment. 

Figure~\ref{fig:Ratios} shows the absolute value of $\mu_B G^B_E/G^B_M$ for each of the outer-ring octet baryons. The large value of this ratio for the $\Sigma^-$ baryon is a result of the choice of normalization; the magnetic moment of the $\Sigma^-$ suggested by the lattice data~\cite{MagFFs} was found to be significantly smaller than the experimental value~\cite{PDG} which is used here. We also note that if the trends displayed for $\mu_B G^B_E/G^B_M$ at the relatively low $Q^2$ values of this study continue to high $Q^2$, zero-crossings of this ratio for the $\Xi^-$ and $\Sigma^-$ baryons seem unlikely.

\section{Conclusion}

We have performed a chiral extrapolation of recent $2+1$-flavor lattice QCD simulation results for the electric Sachs form factors of the outer-ring octet baryons. The simulations used here were performed on a $32^2 \times 64$ volume at six discrete values of the momentum transfer $Q^2$ and six sets of pseudoscalar masses down to $m_\pi\approx 310$~MeV~\cite{MagFFs}.

Independent chiral extrapolations are performed at each value of $Q^2$ for which there are lattice results, using a formalism based on connected heavy baryon chiral perturbation theory. An advantage of this method is that it requires no phenomenological input regarding the $Q^2$ dependence of the form factors.
The proton and neutron form factors extrapolated to the physical pseudoscalar masses agree remarkably well with the experimental determinations, at all values of $Q^2$ considered. This gives a good indication that disconnected quark loop contributions to the nucleon electric form factors are small relative to the uncertainties of this calculation, provided that other systematic effects, such as excited-state contamination, are negligible.

It is notable that the statistical precision of the extrapolated results for the proton electric form factor is in line with parameterizations of experimental results for that quantity. 
In that light, it is particularly important to carefully examine the systematic uncertainties relevant to this work. In particular, we investigate the robustness of the finite-volume corrections used here, which are estimated using leading order chiral perturbation theory.
To this end we present new lattice simulation results at a light pion mass $(m_\pi,m_K)=(220,540)$~MeV and on a larger $48^3\times 96$ lattice. Comparison of these new large-volume points with the extrapolated small-volume results is encouraging. The excellent agreement for the charged baryons in particular indicates that finite-volume effects are well controlled by the estimated finite-volume corrections. It is also clear that the chiral extrapolation performs well; the large-volume results are at a pion mass $\approx 100$~MeV lighter than the lightest of the small-volume results. 

Furthermore, by combining the results of this analysis with those from Ref.~\cite{MagFFs} for the magnetic form factors, we evaluate the ratios $\mu_B G^B_E/G^B_M$ for each of the outer-ring octet baryons. For the proton the results are qualitatively consistent with a linear decrease of this ratio with $Q^2$ as concluded from polarization transfer experiments~\cite{PhysRevC.84.055204,PhysRevC.73.064004,PhysRevC.71.055202}, although the uncertainties are comparatively large.

Finally, we comment that, as was found in Ref.~\cite{MagFFs} for the magnetic form factors, dipole forms in $Q^2$ do not provide a good fit to the lattice simulation results for $G_E$. Dipole-like fits with more general polynomial denominators fare far better.

\section*{Acknowledgements}

The numerical configuration generation was performed using the BQCD lattice QCD program~\cite{Nakamura:2010qh} on the IBM BlueGeneQ using DIRAC 2 resources (EPCC, Edinburgh, UK), the BlueGene P and Q at NIC (J\"ulich, Germany) and the SGI ICE 8200 at HLRN (Berlin-Hannover, Germany). The BlueGene codes were optimized using Bagel~\cite{Boyle:2009vp}. The Chroma software library~\cite{Edwards:2004sx} was used in the data analysis. This work was supported by the EU grants 283286 (HadronPhysics3), 227431 (Hadron Physics2) and by the University of Adelaide and the Australian
Research Council through the ARC Centre of Excellence for Particle Physics at the Terascale and grants FL0992247 (AWT), FT120100821 (RDY), DP140103067 (RDY and JMZ) and FT100100005 (JMZ).

\FloatBarrier
\newpage
\appendix

\section{Chiral perturbation theory extrapolations}
\label{app:ExtrapDetails}

Tables~\ref{tab:app1}, \ref{tab:app2} and \ref{tab:app2} make explicit the values of those chiral coefficients in Eq.~\ref{eq:GEExtrap} which are not given in Ref.~\cite{MagFFs}.
Although the coefficients $\alpha^{Bq}$ and $\overline{\alpha}^{Bq(q')}$ have the same form for both the electric and magnetic form factors, the free parameters $b_D$ ($\sim\mu_D$), $b_f$ ($\sim \mu_F$) and $c_i$ are distinct (and fit separately) for the electric and magnetic cases. The labels `doubly', `singly' and `other' indicate whether the quark $q'$ or $q$ is `doubly- represented', `singly-represented' or not at all represented in the baryon $B$.

\begin{table}[!htbp]
\begin{center}
\begin{ruledtabular}
\begin{tabular}{lcc}
\multicolumn{3}{c}{$\beta_{T}^{B q(\phi)}$} \\
\backslashbox{$m_{\phi}$}{$q$} & $\text{doubly}$ & $\text{singly}$ \\ \hline
$m_{\text{doubly}}+m_{\text{singly}}$ & $2$ & $1$ \\
$m_{\text{singly}}+m_{\text{other}}$ & $\text{}$ & $1$ \\
$m_{\text{doubly}}+m_{\text{other}}$ & $2$ & $\text{}$ \\
$2m_{\text{doubly}}$ & $2$ & $\text{}$ \\
$2m_{\text{singly}}$ & $\text{}$ & $1$ \\
\end{tabular}
\end{ruledtabular}
\end{center}
\caption{Chiral coefficients for tadpole loops, as relevant to Eq.~(\ref{eq:GEExtrap}).}
\label{tab:app1}
\end{table}

\begin{table}[!htbp]
\begin{center}
\begin{ruledtabular}
\begin{tabular}{lccc}
\multicolumn{4}{c}{$\beta_{T}^{\Lambda q(\phi)}$} \\
\backslashbox{$m_{\phi}$}{$q$} & $u$ & $d$ & $s$ \\ \hline
$m_u+m_d$ & $1$ & $1$ & $\text{}$ \\
$m_d+m_s$ & $\text{}$ & $1$ & $1$ \\
$m_u+m_s$ & $1$ & $\text{}$ & $1$ \\
$2m_u$ & $1$ & $\text{}$ & $\text{}$ \\
$2m_d$ & $\text{}$ & $1$ & $\text{}$ \\
$2m_s$ & $\text{}$ & $\text{}$ & $1$ \\
\end{tabular}
\end{ruledtabular}
\end{center}
\caption{Chiral coefficients for tadpole loops involving a $\Lambda$ baryon, as relevant to Eq.~(\ref{eq:GEExtrap}).}
\label{tab:app2}
\end{table}

\begin{table}[!htbp]
\begin{center}
\begin{ruledtabular}
\begin{tabular}{lccc}
\multicolumn{4}{c}{$\beta_{T}^{\Sigma^0 q(\phi)}$} \\
\backslashbox{$m_{\phi}$}{$q$} & $u$ & $d$ & $s$ \\ \hline
$m_u+m_d$ & $1$ & $1$ & $\text{}$ \\
$m_d+m_s$ & $\text{}$ & $1$ & $1$ \\
$m_u+m_s$ & $1$ & $\text{}$ & $1$ \\
$2m_u$ & $1$ & $\text{}$ & $\text{}$ \\
$2m_d$ & $\text{}$ & $1$ & $\text{}$ \\
$2m_s$ & $\text{}$ & $\text{}$ & $1$ \\
\end{tabular}
\end{ruledtabular}
\end{center}
\caption{Chiral coefficients for tadpole loops involving a $\Sigma^0$ baryon, as relevant to Eq.~(\ref{eq:GEExtrap}).}
\label{tab:app3}
\end{table}

\newpage

\section{Fit parameters}
\label{app:fitParams}

\begin{figure}[!h]
\begin{center}
\includegraphics[width=0.48\textwidth]{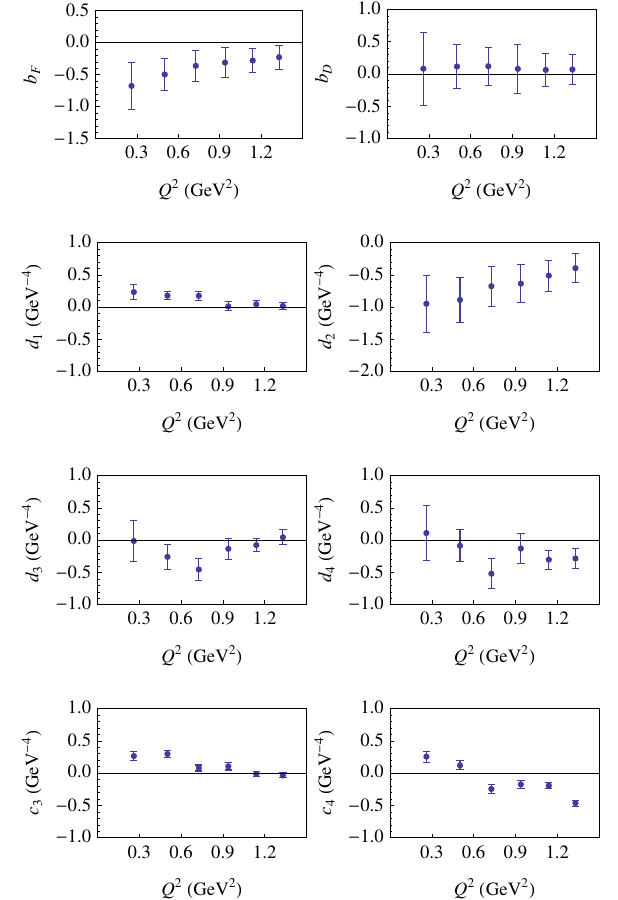}
\caption{$Q^2$ dependence of unrenormalized fit parameters, defined in Eqs.~(\ref{eq:muDF}) and (\ref{eq:LinLag}).}
\label{fig:d2PlotMag}
\end{center}
\end{figure}

This section gives the values of the free parameters determined by the fits to the lattice results. The parameters $b_D$ and $b_F$ are defined in Eq.~(\ref{eq:muDF}), while the $c_i$ appear in Eq.~(\ref{eq:LinLag}). The $d_i$ are relevant linear combinations of the $c_i$:
\begin{align}
d_1 &=c_5-\frac{1}{4}c_{11},& d_2 &= c_6 + c_{11}, \\
d_3 &= c_6 + c_{11}, &d_4&=c_{10}-\frac{5}{2}c_4+c_{12}.
\end{align}
We note that the values of the parameters shown in Fig.~\ref{fig:d2PlotMag} are unrenormalized. They are included merely to illustrate the approximately linear $Q^2$ dependence of the parameters. Recall that the fits at different values of $Q^2$ are independent.

\cleardoublepage
\begin{widetext}
\begin{samepage}

\section{Octet baryon form factors - Figures}
\label{app:FFPics}

Figure~\ref{fig:conn} shows the connected part of the octet baryon electric form factors, extrapolated to the physical pseudoscalar masses. The fits shown are those used in Sec.~\ref{sec:radii} to extract the electric radii.
\vspace{1cm}
\begin{figure*}[!h]
\begin{center}
\subfigure[]{
\includegraphics[width=0.47\textwidth]{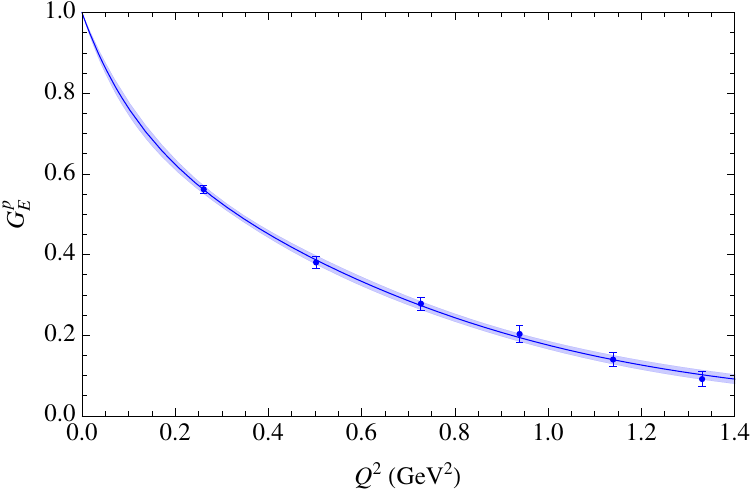}
\label{fig:Ep}
}
\subfigure[]{
\includegraphics[width=0.47\textwidth]{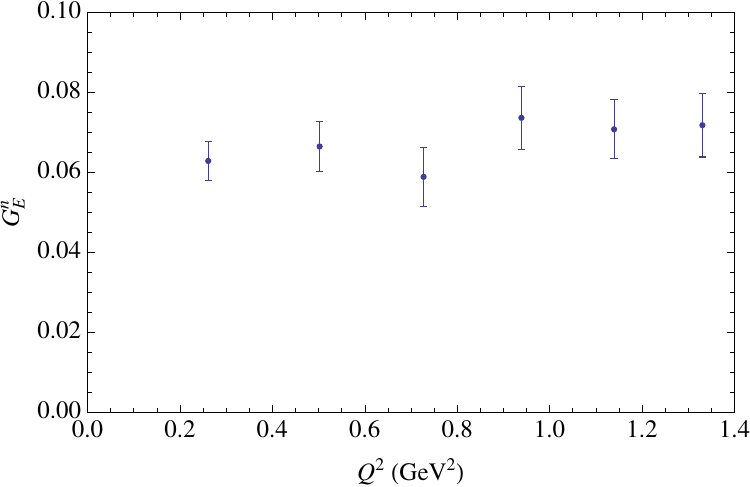}
\label{fig:En}
}
\subfigure[]{
\includegraphics[width=0.47\textwidth]{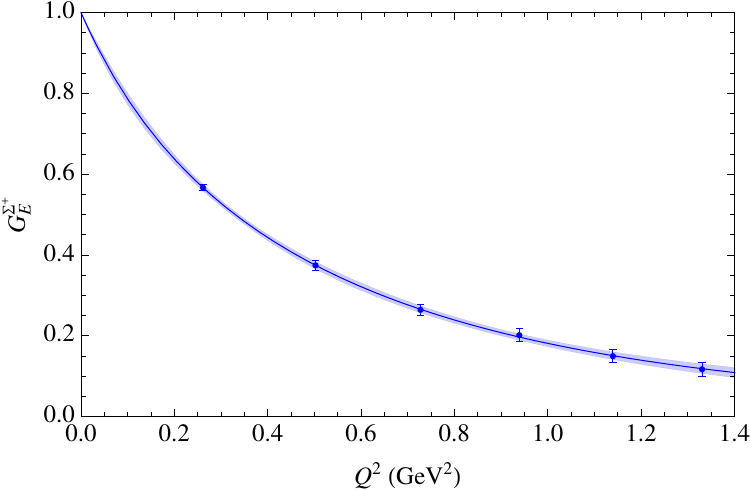}
\label{fig:MMSp}
}
\subfigure[]{
\includegraphics[width=0.47\textwidth]{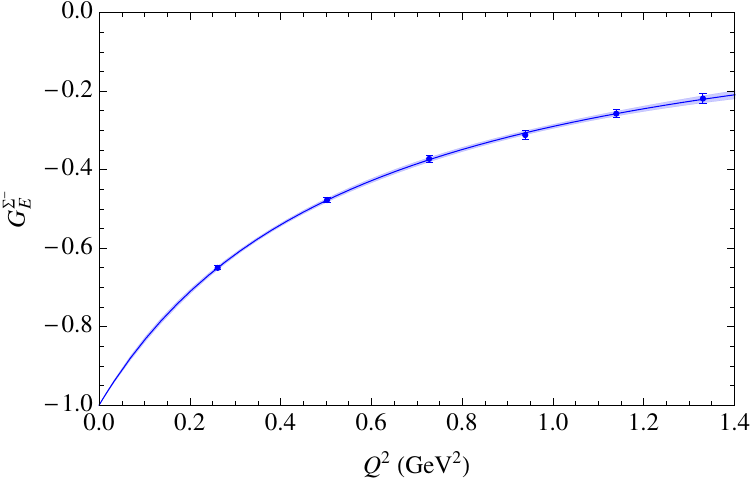}
\label{fig:MMSm}
}
\subfigure[]{
\includegraphics[width=0.47\textwidth]{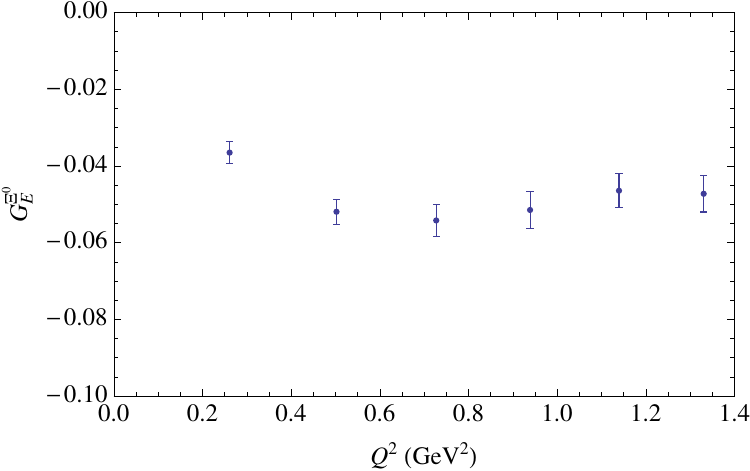}
\label{fig:MMX0}
}
\subfigure[]{
\includegraphics[width=0.47\textwidth]{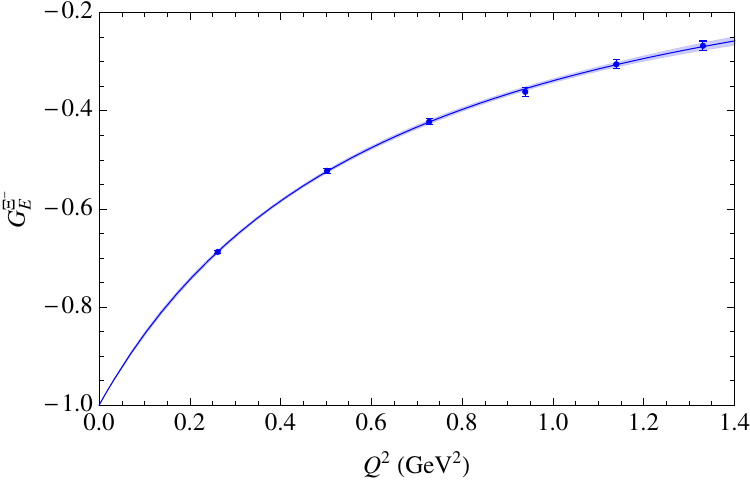}
\label{fig:MMXm}
}
\caption{Connected part of the octet baryon electric form factors. Lines shown for the charged baryons correspond to dipole-like fits (Eq.~(\ref{eq:diplike})).}
\label{fig:conn}
\end{center}
\end{figure*}

\cleardoublepage
\end{samepage}
\end{widetext}

\pagebreak

\section{Raw lattice simulation results}
\label{app:RawLatt}

\begin{figure}[]
\begin{center}
\subfigure[Doubly-represented quark contributions.]{
\includegraphics[width=0.48\textwidth]{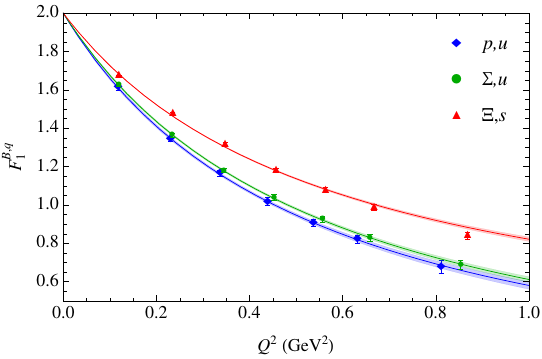}
\label{fig:F1doubly}
}
\subfigure[Singly-represented quark contributions.]{
\includegraphics[width=0.48\textwidth]{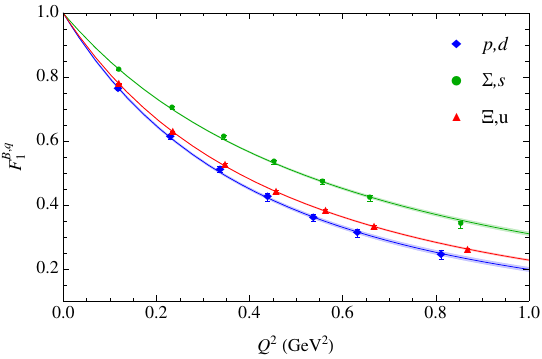}
\label{fig:F1singly}
}
\caption{Quark contributions to the Dirac form factor $F_1$ of the hyperons for the $L^3\times T=48^3 \times 96$ simulation detailed in Table~\ref{tab:SimDetailsNew}. The charges of the relevant quarks have been set to unity. Lines correspond to dipole-like fits (Eq.~(\ref{eq:F1Fit})).}
\label{fig:F1}
\end{center}
\end{figure}

\begin{figure}[]
\begin{center}
\subfigure[Doubly-represented quark contributions.]{
\includegraphics[width=0.48\textwidth]{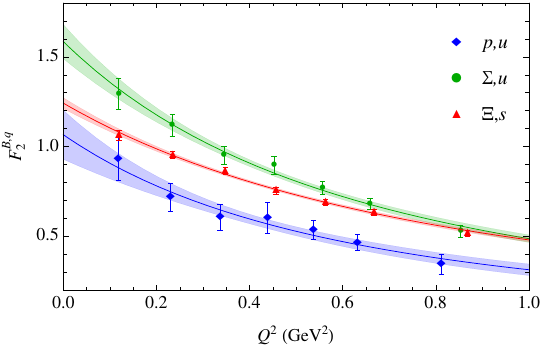}
\label{fig:F2doubly}
}
\subfigure[Singly-represented quark contributions.]{
\includegraphics[width=0.48\textwidth]{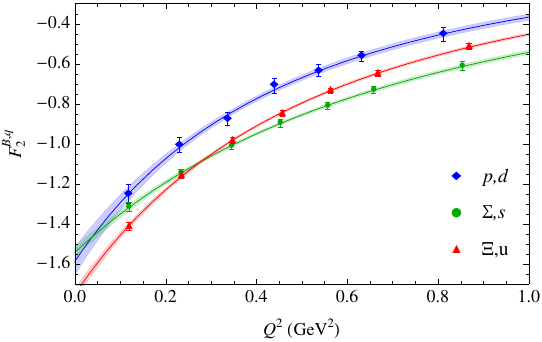}
\label{fig:F2singly}
}
\caption{Quark contributions to the Pauli form factor $F_2$ of the hyperons for the $L^3\times T=48^3 \times 96$ simulation detailed in Table~\ref{tab:SimDetailsNew}. The charges of the relevant quarks have been set to unity. Lines correspond to dipole fits (Eq.~(\ref{eq:F2Fit})).}
\label{fig:F2}
\end{center}
\end{figure}

Here we present raw lattice results for the $L^3\times T=48^3 \times 96$ simulation detailed in Table~\ref{tab:SimDetailsNew}. Similar details for the other lattice simulations used in this study may be found in Ref.~\cite{MagFFs}. We point out that no finite-volume correction or chiral extrapolation has been applied to these numbers.
Tables~\ref{tab:nucres}, \ref{tab:sigres}, \ref{tab:xires} give results for $F_1$ and $F_2$. Figures~\ref{fig:F1} and \ref{fig:F2} show the results pictorially.
The fits shown in those figures use the 2-parameter ans\"atze: 
%
\begin{align}
\label{eq:F1Fit}
F_1(Q^2) & = \frac{F_1(0)}{1 + c_{12}Q^2 + c_{14}Q^4}, \\ \label{eq:F2Fit}
F_2(Q^2) & = \frac{F_2(0)}{(1 + c_{22}Q^2)^2},
\end{align}
where the $c_{ij}$ and the anomalous magnetic moment $F_2^{B,q}(0) =\kappa^{B,q}$ are fit parameters, while $F_1(0)$ is fixed by charge conservation. As we consider quarks of unit charge, $F_1(0)=2,1$ for the doubly and singly represented quarks respectively. Dirac and Pauli mean-squared charge radii extracted from these fits are given in Table~\ref{tab:radii}.

\begin{table}[p]
\begin{ruledtabular}
\begin{tabular}{cD{.}{.}{1}D{.}{.}{1}D{.}{.}{1}}
 \multicolumn{1}{c}{$B,q$} & \multicolumn{1}{c}{ $\langle r^2 \rangle_1^{B,q}$} & \multicolumn{1}{c}{ $\langle r^2 \rangle_2^{B,q}$} & \multicolumn{1}{c}{$\kappa^{B,q}$}  \\ \hline
$p,u$ & $\text{0.467 (16)}$ & $\text{0.391 (91)}$ & $\text{1.06 (13)}$ \\
$p,d$ &$\text{0.558 (19)}$ & $\text{0.502 (39)}$ & $\text{-1.582 (69)}$ \\
$\Sigma,u$ & $\text{0.441 (10)}$ & $\text{0.374 (40)}$ & $\text{1.580 (95)}$ \\
$\Sigma,s$ & $\text{0.4008 (69)}$ & $\text{0.319 (14)}$ & $\text{-1.536 (28)}$ \\
$\Xi,s$ & $\text{0.3732 (35)}$ & $\text{0.283 (16)}$ & $\text{1.238 (28)}$ \\
$\Xi,u$ & $\text{0.5208 (69)}$ & $\text{0.450 (13)}$ & $\text{-1.744 (28)}$ \\
\end{tabular}
\end{ruledtabular}
\caption{Dirac and Pauli mean-squared charge radii and anomalous magnetic moments for the $L^3\times T=48^3 \times 96$ lattice simulation results, extracted from dipole-like fits (see Eqs.~(\ref{eq:F1Fit}) and (\ref{eq:F2Fit})).}
\label{tab:radii}
\end{table}


\begin{table*}[p]
\begin{ruledtabular}
\begin{tabular}{ccD{.}{.}{-1}D{.}{.}{1}D{.}{.}{1}D{.}{.}{1}D{.}{.}{1}}
$m_\pi$ (MeV) & $m_K$ (MeV) & \multicolumn{1}{c}{$Q^2$ (GeV$^2$)} & \multicolumn{1}{c}{$F_1^{p,u}$} & \multicolumn{1}{c}{$F_1^{p,d}$} & \multicolumn{1}{c}{$F_2^{p,u}$} & \multicolumn{1}{c}{$F_2^{p,d}$} \\ \hline
$220$ & $540$ & $0.12$ & $\text{1.612 (12)}$ & $\text{0.7631 (50)}$ & $\text{0.93 (12)}$ & $\text{-1.251 (46)}$ \\
$\text{}$ & $\text{}$ & $0.23$ & $\text{1.342 (12)}$ & $\text{0.6122 (75)}$ & $\text{0.717 (76)}$ & $\text{-1.006 (36)}$ \\
$\text{}$ & $\text{}$ & $0.34$ & $\text{1.165 (16)}$ & $\text{0.5103 (93)}$ & $\text{0.606 (74)}$ & $\text{-0.877 (32)}$ \\
$\text{}$ & $\text{}$ & $0.44$ & $\text{1.016 (21)}$ & $\text{0.424 (12)}$ & $\text{0.604 (85)}$ & $\text{-0.709 (38)}$ \\
$\text{}$ & $\text{}$ & $0.54$ & $\text{0.906 (18)}$ & $\text{0.359 (10)}$ & $\text{0.534 (53)}$ & $\text{-0.635 (29)}$ \\
$\text{}$ & $\text{}$ & $0.63$ & $\text{0.822 (20)}$ & $\text{0.311 (11)}$ & $\text{0.465 (45)}$ & $\text{-0.563 (26)}$ \\
$\text{}$ & $\text{}$ & $0.81$ & $\text{0.678 (36)}$ & $\text{0.244 (15)}$ & $\text{0.345 (54)}$ & $\text{-0.452 (36)}$ \\
\end{tabular}
\end{ruledtabular}
\caption{Raw lattice simulation results for the nucleon for the $L^3\times T=48^3 \times 96$ simulation detailed in Table~\ref{tab:SimDetailsNew}.}
\label{tab:nucres}
\end{table*}

\begin{table*}[p]
\begin{ruledtabular}
\begin{tabular}{ccD{.}{.}{-1}D{.}{.}{1}D{.}{.}{1}D{.}{.}{1}D{.}{.}{1}}
$m_\pi$ (MeV) & $m_K$ (MeV) & \multicolumn{1}{c}{$Q^2$ (GeV$^2$)} & \multicolumn{1}{c}{$F_1^{\Sigma,u}$} & \multicolumn{1}{c}{$F_1^{\Sigma,s}$} & \multicolumn{1}{c}{$F_2^{\Sigma,u}$} & \multicolumn{1}{c}{$F_2^{\Sigma,s}$} \\ \hline
$220$ & $540$ & $0.12$ & $\text{1.6270 (77)}$ & $\text{0.8219 (21)}$ & $\text{1.294 (88)}$ & $\text{-1.314 (21)}$ \\
$\text{}$ & $\text{}$ & $0.23$ & $\text{1.3616 (92)}$ & $\text{0.7040 (38)}$ & $\text{1.120 (60)}$ & $\text{-1.147 (18)}$ \\
$\text{}$ & $\text{}$ & $0.35$ & $\text{1.178 (10)}$ & $\text{0.6109 (52)}$ & $\text{0.952 (49)}$ & $\text{-1.010 (17)}$ \\
$\text{}$ & $\text{}$ & $0.45$ & $\text{1.037 (15)}$ & $\text{0.5335 (74)}$ & $\text{0.896 (49)}$ & $\text{-0.898 (19)}$ \\
$\text{}$ & $\text{}$ & $0.56$ & $\text{0.924 (16)}$ & $\text{0.4723 (80)}$ & $\text{0.772 (35)}$ & $\text{-0.810 (18)}$ \\
$\text{}$ & $\text{}$ & $0.66$ & $\text{0.829 (16)}$ & $\text{0.4202 (87)}$ & $\text{0.681 (30)}$ & $\text{-0.731 (19)}$ \\
$\text{}$ & $\text{}$ & $0.85$ & $\text{0.687 (25)}$ & $\text{0.338 (11)}$ & $\text{0.530 (34)}$ & $\text{-0.610 (21)}$ \\
\end{tabular}
\end{ruledtabular}
\caption{Raw lattice simulation results for the sigma baryon for the $L^3\times T=48^3 \times 96$ simulation detailed in Table~\ref{tab:SimDetailsNew}.}
\label{tab:sigres}
\end{table*}

\begin{table*}[p]
\begin{ruledtabular}
\begin{tabular}{ccD{.}{.}{-1}D{.}{.}{1}D{.}{.}{1}D{.}{.}{1}D{.}{.}{1}}
$m_\pi$ (MeV) & $m_K$ (MeV) & \multicolumn{1}{c}{$Q^2$ (GeV$^2$)} & \multicolumn{1}{c}{$F_1^{\Xi,s}$} & \multicolumn{1}{c}{$F_1^{\Xi,u}$} & \multicolumn{1}{c}{$F_2^{\Xi,s}$} & \multicolumn{1}{c}{$F_2^{\Xi,u}$} \\ \hline
$220$ & $540$ & $0.12$ & $\text{1.6759 (21)}$ & $\text{0.7779 (20)}$ & $\text{1.062 (26)}$ & $\text{-1.410 (19)}$ \\
$\text{}$ & $\text{}$ & $0.24$ & $\text{1.4772 (47)}$ & $\text{0.6288 (27)}$ & $\text{0.955 (21)}$ & $\text{-1.155 (16)}$ \\
$\text{}$ & $\text{}$ & $0.35$ & $\text{1.3183 (71)}$ & $\text{0.5251 (33)}$ & $\text{0.862 (19)}$ & $\text{-0.982 (16)}$ \\
$\text{}$ & $\text{}$ & $0.46$ & $\text{1.1835 (94)}$ & $\text{0.4400 (47)}$ & $\text{0.756 (19)}$ & $\text{-0.848 (14)}$ \\
$\text{}$ & $\text{}$ & $0.56$ & $\text{1.079 (11)}$ & $\text{0.3800 (46)}$ & $\text{0.691 (17)}$ & $\text{-0.734 (13)}$ \\
$\text{}$ & $\text{}$ & $0.67$ & $\text{0.987 (13)}$ & $\text{0.3310 (48)}$ & $\text{0.636 (17)}$ & $\text{-0.648 (13)}$ \\
$\text{}$ & $\text{}$ & $0.87$ & $\text{0.840 (18)}$ & $\text{0.2594 (61)}$ & $\text{0.518 (17)}$ & $\text{-0.512 (14)}$ \\
\end{tabular}
\end{ruledtabular}
\caption{Raw lattice simulation results for the cascade baryon for the $L^3\times T=48^3 \times 96$ simulation detailed in Table~\ref{tab:SimDetailsNew}.}
\label{tab:xires}
\end{table*}

\FloatBarrier

\bibliography{ElecFFBib}

\end{document}